\documentclass[11pt,a4paper]{article}
\pdfoutput=1

\usepackage{jheppub}
\usepackage{latexsym}
\usepackage{multirow}
\usepackage{color}
\usepackage[usenames,dvipsnames,table]{xcolor}
\usepackage{graphicx}
\usepackage{epsfig}  
\usepackage{epsf}    
\usepackage{dcolumn}
\usepackage{bm}
\usepackage{dcolumn}
\usepackage{textcomp}
\usepackage{float}
\usepackage{subfig}
\usepackage{hypcap}
\usepackage[]{hyperref}
\usepackage{makecell}
\usepackage{color}
\usepackage{pifont}
\usepackage{appendix}
\usepackage{url}

\hypersetup{
  bookmarks=true,         
  unicode=false,          
  pdftoolbar=true,        
 pdfmenubar=true,        
 pdffitwindow=true,     
 pdfstartview={FitH},    
 pdfsubject={Neutrino Oscillations Phenomenology},   
 pdfnewwindow=true,      
 pdfcreator={RevTeX},
 colorlinks=true,       
 linkcolor=red,          
 citecolor=blue,        
 filecolor=black,      
 urlcolor=blue,           
  }

\newcommand{\be}{\begin{equation}}
\newcommand{\ee}{\end{equation}}
\newcommand{\ba}{\begin{eqnarray}}
\newcommand{\ea}{\end{eqnarray}}

\makeatletter
\renewcommand{\fnum@table}{\textbf{\tablename~\thetable}}
\renewcommand{\fnum@figure}{\textbf{\figurename~\thefigure}}
\makeatother

\preprint{TIFR/TH/17-06}

\title{Neutrino mixing and $R_K$ anomaly in $U(1)_X$ models: \\
a bottom-up approach} 

\author{Disha Bhatia,} 
\author{Sabyasachi Chakraborty,}
\author{Amol Dighe} 

\affiliation{Tata Institute of Fundamental Research, Mumbai 400005, India}

\emailAdd{disha@theory.tifr.res.in}
\emailAdd{sabya@theory.tifr.res.in}
\emailAdd{amol@theory.tifr.res.in}

\abstract
{
We identify a class of $U(1)_X$ models which can explain the 
$R_K$ anomaly and the neutrino mixing pattern, by using a bottom-up approach.
The different $X$-charges of lepton generations account for the
lepton universality violation required to explain $R_K$.
In addition to the three right-handed neutrinos needed for the Type-I seesaw
mechanism, these minimal models only introduce an additional doublet Higgs and
a singlet scalar. 
While the former helps in reproducing the quark mixing structure, the latter 
gives masses to neutrinos and the new gauge boson $Z^\prime$.
Our bottom-up approach determines the $X$-charges of all particles using
theoretical consistency and experimental constraints.
We find the parameter space allowed by the constraints from neutral meson
mixing, rare $b \to s$ decays and direct collider searches for $Z^\prime$.
Such a $Z^\prime$ may be observable at the ongoing run of the 
Large Hadron Collider with a few hundred fb$^{-1}$ of integrated luminosity.
}


\begin{document}
\maketitle
\flushbottom

\newpage

\section{Introduction}
\label{intro}

We live in an era enriched with many experimental breakthroughs and results. 
The recent discovery of the 125 GeV Higgs boson 
at the Large Hadron Collider (LHC) has marked the completion of the Standard Model (SM).
However, physics beyond the SM (BSM) is certain to exist, and would be needed to explain 
observations like neutrino masses and mixings, matter-antimatter asymmetry, and dark matter.   
Although the direct searches performed by the two LHC-based experiments, viz. ATLAS and CMS, 
have not yet found any new particle, indirect hints of new physics (NP) may still be hidden 
in the data. 

Recently, the LHCb collaboration has reported some indirect hints of BSM physics in 
the $b \to s \ell \ell$ flavour observables. Major among these are the
measurements of $R_K$, defined as the ratio of branching fractions of $B^+ \to K^+ \mu^+ \mu^-$ 
and $B^+ \to K^+ e^+ e^-$ in the low dilepton mass-squared bin~\cite{RKexp}:
\be
R_K \equiv \left. 
\frac{{\rm BR}(B \to K \mu \mu)}{{\rm BR}(B \to K e e)} \right|_{q^2 = 1-6 ~{\rm GeV}^2} \; ,
\ee
and the angular 
observable $P_5^\prime$~\cite{DescotesGenon:2012zf} in the decays of the $B$ mesons in $B\rightarrow K^* \mu \mu$~\cite{P51,P52}.
The BELLE collaboration has also reported an anomaly in $P_5^\prime$~\cite{Wehle:2016yoi}
which is compatible with the one observed in~\cite{P51,P52}.
The branching ratio measurements of $B \to K^* \mu \mu$~\cite{btok*} and 
$B \to \phi \mu \mu$~\cite{btophi} also show slight deviations from the SM predictions. 
While the latter anomalies could be accounted for by form factor uncertainties, 
the $R_K$ measurement should be free from strong interaction effects, since the form 
factors cancel in the ratio. Therefore, if the $R_K$ anomaly is confirmed, it would signal 
a clear lepton flavour universality violation~\cite{rk_sm_qcd,rk_sm_qed}.

The anomalies in $R_K$ and $P_5'$ measurements can be addressed by invoking
additional NP contributions to some of the Wilson coefficients $C_i(\mu)$ appearing in 
the effective Hamiltonian for $b \to s \ell \ell$. In the SM, the effective
Hamiltonian for this process is~\cite{buchalla}
\begin{eqnarray}
{\mathcal H}_{\rm eff} & = & -\frac{4 G_{F}}{\sqrt{2}} V_{tb}V^{*}_{ts} \times \nonumber \\
& & \left(  \sum_{i=1,6} C_i {\mathcal O}_{i} +C_{7\gamma} \mathcal{O}_{7\gamma}  
+ C_{8G} {\mathcal O}_{8G}  
+  \sum_{i=9,10} C_i \mathcal{O}_{i}  
+ \sum_{i=S,P} C_{i}^{(\prime)} \mathcal{O}_{i}^{(\prime)} \right) \;,
\label{heff}
\end{eqnarray}
where $\mathcal{O}_i$'s are the effective operators, and $^\prime$ indicates currents
with opposite chirality. The values of $C_i(m_b)$ have been calculated in \cite{wilson_mb}.
At the leading order, the additional NP contributions may contribute to the operators 
which are already present in the SM:
 \begin{eqnarray}
 \mathcal{O}_{7\gamma} &=&  \frac{e}{16 \pi^2} m_b \left( \bar{s} \sigma_{\mu \nu} P_R b \right) F^{\mu \nu}\;,~ 
\mathcal{O}_9 =  \frac{\alpha_e}{4 \pi}\left[ \bar{s}\gamma_{\mu} P_L b\right] \left[ \bar{\ell}\gamma^{\mu}  \ell\right]\;,~\nonumber \\
\mathcal{O}_{10} &=& \frac{\alpha_e}{4 \pi}\left[ \bar{s}\gamma_{\mu} P_L b\right] \left[ \bar{\ell}\gamma^{\mu}  \gamma_5 \ell\right]\;,
\label{eqn:SM}
\end{eqnarray}
or may enhance the effects of the operators whose contributions are normally suppressed by the 
lepton mass in the SM:
\begin{eqnarray}
\mathcal{O}_S = \frac{\alpha_e}{4 \pi} \left[ \bar{s} P_R b\right] \left[ \bar{\ell} \ell\right] \;, &\quad &
\mathcal{O}_P =  \frac{\alpha_e}{4 \pi} \left[ \bar{s} P_R b\right] \left[ \bar{\ell} \gamma_5 \ell\right], \nonumber \\
\mathcal{O}_S^{\prime}  =\frac{\alpha_e}{4 \pi}    \left[ \bar{s} P_L b\right] \left[ \bar{\ell} \ell\right] \;, &\quad &
\mathcal{O}_P^{\prime}  =  \frac{\alpha_e}{4 \pi}  \left[ \bar{s} P_L b\right] \left[ \bar{\ell} \gamma_5 \ell\right] \; ,
\label{eqn:sup}
\end{eqnarray} 
or may generate new operators which are absent in SM \cite{Hiller:2003js}:
\begin{eqnarray}
\mathcal{O}^{\prime}_{7\gamma}&=& \frac{e}{16 \pi^2} m_b \left( \bar{s} \sigma_{\mu \nu} P_L b \right) F^{\mu \nu}\;, ~
 \mathcal{O}^{\prime}_9 = \frac{\alpha_e}{4 \pi}\left[ \bar{s}\gamma_{\mu} P_R b\right] 
\left[ \bar{\ell}\gamma^{\mu} \ell\right] \;,\nonumber \\
\mathcal{O}_{10}^{\prime} &=& \frac{\alpha_e}{4 \pi}\left[ \bar{s}\gamma_{\mu} P_R b\right]
 \left[ \bar{\ell}\gamma^{\mu}  \gamma_5 \ell\right] \; .
\label{eqn:BSM}
\end{eqnarray}

Simultaneous explanation of the $R_K$ and $P_5^\prime$ anomalies is possible if the NP effects 
are present in ${\mathcal O}_9, {\mathcal O}_9^\prime, {\mathcal O}_{10}$ or ${\mathcal O}_{10}^\prime$ 
operators~\cite{rk_eft}. The global fits 
~\cite{Descotes-Genon:2013wba,globalfit1,globalfit2,globalfit3,globalfit4} prefer NP effects in 
$\mathcal{O}_9^{\mu}$, i.e. additional contributions to $C_9^{\mu}$.
Since the observed value of $R_K {\rm (obs)} = 0.745 ^{+0.090}_{-0.074} \pm 0.036$ \cite{RKexp}
is less than the SM prediction, which gives $R_K$ to be unity within an accuracy of 1\%
~\cite{rk_sm_qcd,rk_sm_qed},
the new physics contribution must interfere destructively with the SM,
i.e. opposite to that of $C_9^{\rm SM}(m_b) = 4.2$~\cite{wilson_mb}.  
This indicates that the sign of $C_9^{\text{NP},\mu}$
is negative. The best-fit value of $C_9^{\text{NP},\mu}$ is $\approx -1$ 
~\cite{Descotes-Genon:2013wba,globalfit1,globalfit2,globalfit3,globalfit4,rk_eft}.
In addition $C_9^{\text{NP},\mu} = -C_{10}^{\text{NP},\mu}$ also gives a good fit to data
~\cite{globalfit1,globalfit2,globalfit3,globalfit4}.
Motivated by these results, many explanations of the anomaly using $Z^\prime$
~\cite{Gauld:2013qba,Glashow:2014iga,Bhattacharya:2014wla, Crivellin:2015mga,
horizontalpaper,Crivellin:2015era,Celis:2015ara,Sierra:2015fma,Belanger:2015nma,Gripaios:2015gra,
Allanach:2015gkd,Fuyuto:2015gmk,Chiang:2016qov,Boucenna:2016wpr,Boucenna:2016qad,Celis:2016ayl,
Altmannshofer:2016jzy,
Bhattacharya:2016mcc,Crivellin:2016ejn,Becirevic:2016zri,GarciaGarcia:2016nvr} and leptoquark~\cite{rk_eft,Biswas:2014gga,Gripaios:2014tna,Sahoo:2015wya,Becirevic:2015asa,
Alonso:2015sja,Calibbi:2015kma,
Huang:2015vpt,Pas:2015hca,Bauer:2015knc,Fajfer:2015ycq,Barbieri:2015yvd,
Sahoo:2015pzk,
Dorsner:2016wpm,Sahoo:2016nvx,Das:2016vkr,Chen:2016dip,Becirevic:2016oho,Becirevic:2016yqi,Bhattacharya:2016mcc,Sahoo:2016pet,
Barbieri:2016las,Cox:2016epl} 
models have been given in the literature.

Since the flavour anomalies mentioned above mostly involve muons, and there is no 
clear hint of new physics effects in the electron sector apart from $R_K$ measurement, 
most of the analysis have been performed assuming new physics effects in muons only. 
However, NP contributions in the electron sector, $C_9^{\text{NP},e}$, of the same order as those 
in the muon sector, are still consistent with all $b \to s$ measurements within 
2$\sigma$~\cite{globalfit1,globalfit2,globalfit3,globalfit4}.
The comparisons among two dimensional global fits also prefer 
($C_9^{\text{NP},e}$, $C_{9}^{\text{NP},\mu}$) 
over other combinations like ($C_9^{\text{NP},\mu}$, $C_{10}^{\text{NP},\mu}$) and 
($C_9^{\text{NP},\mu}$, ${C_{9}^{\prime~\text{NP},\mu}}$), with the best fit point 
favouring dominant contributions to $C_9^{\text{NP},\mu}$~\cite{globalfit4}.

In this work, we build our analysis around the choice where NP contributes via the 
$\mathcal{O}_9$ operator. We allow both $C^{\text{NP},e}_9$ and $C^{\text{NP},\mu}_9$ to be present.
Since these two contributions have to be different, the NP must violate
lepton flavour universality. This may be implemented in a minimalistic way through an abelian 
symmetry $U(1)_X$, under which the leptons have different charges. 
In particular, greater NP contribution to $C_9^{\text{NP},\mu}$ than $C_9^{\text{NP},e}$
may be achieved by a higher magnitude of the $X$-charge for muons than for electrons. 
Substantial NP contributions to the flavour anomalies also require  
tree-level flavour-changing neutral currents (FCNC) in the quark sector.
These can be implemented through different $X$-charges for the quark 
generations as well, which should still allow for quark mixing, and
be consistent with the flavour physics data. 

A horizontal $U(1)_X$ symmetry in the lepton sector would also determine the possible 
textures in the mass matrix of the right-handed neutrinos. In turn, the mixing pattern 
of the left-handed neutrinos~\cite{Capozzi:2016rtj,Esteban:2016qun}
will be affected through the Type-I seesaw mechanism.
The possible textures of the right-handed neutrino mass matrix and the lepton flavour 
universality violation required for the flavour anomalies can thus have a common origin. 
Scenarios like an $L_\mu-L_\tau$ symmetry with $X$-charges given to the SM quarks
\cite{horizontalpaper,Altmannshofer:2016jzy} 
or additional vector-like quarks \cite{Crivellin:2015mga,Fuyuto:2015gmk},
have been considered in the literature in this context. 
Other models with $Z^\prime$ also have their own $X$-charge assignments 
\cite{Crivellin:2015era,Celis:2015ara,Belanger:2015nma,Allanach:2015gkd,Celis:2016ayl}, 
however their connection with the neutrino mass matrix has not been explored. 
We build our model in the bottom-up approach, where we do not assign the $X$-charges a priori, 
but look for the $X$-charge assignments that satisfy the data in the quark and lepton sectors.
As a guiding principle, we introduce a minimal number of additional particles, and 
ensure that the model is free of any gauge anomalies. Finally, we identify the horizontal
symmetries that are compatible with the observed neutrino mixing pattern, and at the same 
time are able to generate $C_9^{\text{NP},e}$ and $C_{9}^{\text{NP},\mu}$ that explain the 
flavour anomalies.

The paper is organized as follows. 
In section~\ref{section:construction}, we describe the construction of the $U(1)_X$ models
from a bottom-up approach. 
In section~\ref{section:exp_constraints}, we explore the allowed ranges of the parameters
that are consistent with the experimental constraints like neutral meson mixings, rare 
$B$ decays, and direct collider searches for $Z^\prime$.
In section~\ref{section:predictions}, we present the predictions for the CP-violating 
phases in the lepton sectors for specific horizontal symmetries, and project the reach
of the LHC for detecting the corresponding $Z^\prime$.
In section~\ref{section:conclusions}, we summarize our results and present our concluding remarks.
Further in appendix~\ref{section:zprime}, we present the generation of the $Z^\prime$ mass and 
$Z$--$Z^\prime$ mixing. In appendix~\ref{section:FCNC}, we discuss the constraints on the flavour changing neutral
interactions from the scalar sector and in appendix~\ref{section:nunu}, we calculate the effects of our model
on $b \to s \nu \nu$ transitions.     

\section{Constructing the $U(1)_X$ class of models}
\label{section:construction}

We construct a class of models wherein, in addition to the SM fields, we also have three
right-handed neutrinos that would be instrumental in giving mass to the
left-handed neutrinos through the seesaw mechanism. 
We extend the SM gauge symmetry group by an additional symmetry, $U(1)_X$,  which 
corresponds to an additional gauge boson, $Z^\prime$, with mass $M_{Z^\prime}$ and 
gauge coupling $g_{Z^\prime}$. 
To start with, we denote the $X$-charge for a SM field $i$ by $X_{i}$.  
In this section, we shall determine the values of $X_{i}$'s in a bottom-up approach. 

\subsection{Preliminary constraints on the $X$-charges}
\label{sec:prelim}

Since we wish to build up the model by introducing NP effects only in the $\mathcal{O}_9$
operator, we have to make sure that the NP contribution to all the other operators listed in 
eqs.~(\ref{eqn:SM}),~(\ref{eqn:sup}), and (\ref{eqn:BSM}) should vanish.
We first consider the interactions of $Z^\prime$ with charged leptons, $\ell$, in the mass basis:  
\begin{equation}
\mathcal{L}^\ell_{Z^\prime} = g_{Z^\prime} \, \overline{\ell_L}  \, \gamma^\mu V^\dagger_{\ell_{L}} \,
{\mathcal X}_{\ell_{L}} \, V_{\ell_{L}} \, \ell_L \, Z^\prime_\mu
+ g_{Z^\prime} \, \overline{\ell_R} \, \gamma^\mu V^\dagger_{\ell_{R}} \,
{\mathcal X}_{\ell_{R}} \, V_{\ell_{R}} \, \ell_R \,Z^\prime_\mu \;,
\label{eq:lepton-Z}
\end{equation}
where ${\mathcal X}_{\ell_{L}} = \text{diag}\left( X_{e_L}, X_{\mu_L}, X_{\tau_L}\right) $ and 
${\mathcal X}_{\ell_{R}} = \text{diag}\left( X_{e_{R}}, X_{\mu_{R}}, X_{\tau_{R}}\right)$, while
$V_{\ell_L}$ and $V_{\ell_{R}}$ are the rotation matrices diagonalizing the Yukawa matrix 
for charged leptons. Note that the $SU(2)_L$ gauge invariance of the SM ensures 
${\mathcal X}_{\ell_{L}} = {\mathcal X}_{{\nu_\ell}_{L}}$.

The Lagrangian in eq.~(\ref{eq:lepton-Z}) may be rewritten as
\ba
\mathcal{L}^\ell_{Z^\prime} & = & \frac{1}{2} \, g_{Z^\prime} \, \overline{\ell} \, \gamma^\mu  
\left( V^\dagger_{\ell_{L}} {\mathcal X}_{\ell_{L}} V_{\ell_{L}} 
  + V^\dagger_{\ell_{R}} {\mathcal X}_{\ell_{R}} V_{\ell_{R}} \right) \ell \,Z^\prime_\mu \nonumber \\
& &  - \frac{1}{2} \, g_{Z^\prime} \, \overline{\ell} \, \gamma^\mu \gamma_5 
\left( V^\dagger_{\ell_{L}} {\mathcal X}_{\ell_{L}} V_{\ell_{L}} 
  - V^\dagger_{\ell_{R}} {\mathcal X}_{\ell_{R}} V_{\ell_{R}} \right) \ell \,Z^\prime_\mu \;.
\label{equation:lepLag}
\ea
The second term in eq.~(\ref{equation:lepLag}) would contribute to $\mathcal{O}_{10}$ and
$\mathcal{O}_{10}^\prime$. Since we do not desire such a contribution, we require
\begin{equation} 
V^\dagger_{\ell_{L}} {\mathcal X}_{\ell_{L}} V_{\ell_{L}} = V^\dagger_{\ell_{R}} {\mathcal X}_{\ell_{R}} V_{\ell_{R}}\;.
\label{eq:VXV}
\end{equation}
  
A straight forward solution to the eq.~(\ref{eq:VXV}) yields $V_{\ell_{L}} = I$ and $V_{\ell_{R}} = I$
and further ${\mathcal X}_{\ell_{L}} = {\mathcal X}_{\ell_{R}}$. In such a case a non-zero Yukawa matrix 
would need the Higgs field, $\Phi$, to be a singlet under $U(1)_X$. Note that with unequal vector-like charge 
assignments in the lepton sector, the Yukawa matrix will naturally be diagonal. This therefore is a 
minimal and consistent solution and we proceed with this in our analysis.

Now we turn to the $Z^\prime$ interactions with the $d$-type quarks:
\begin{equation}
\mathcal{L}^d_{Z^\prime} = g_{Z^\prime} \, \overline{d_L}  \, \gamma^\mu  V^\dagger_{d_{L}} \,
{\mathcal X}_{d_{L}} \, V_{d_{L}} d_L \,  Z^\prime_\mu+ 
g_{Z^\prime} \, \overline{d_R} \, \gamma^\mu  V^\dagger_{d_{R}} \,
{\mathcal X}_{d_{R}} \, V_{d_{R}} \, d_R \,Z^\prime_\mu \;,
\label{equation:quarkLag}
\end{equation}    
where ${\mathcal X}_{d_{L}} = \text{diag}\left( X_{d_L}, X_{s_L}, X_{b_L}\right)$, 
${\mathcal X}_{d_{R}} = \text{diag}\left( X_{d_{R}}, X_{s_{R}}, X_{b_{R}}\right)$,
while $V_{d_{L}}$ and $V_{d_{R}}$ are the rotation matrices which diagonalize the Yukawa
matrix for $d$-type quarks.  
Note that the $SU(2)_L$ gauge invariance of the SM ensures ${\mathcal X}_{d_{L}} = {\mathcal X}_{u_{L}}$.

Substantial NP effects require the $X$-charges 
to be non-universal, thereby generating both  $\overline{b_L} \gamma^\mu s_L Z^{\prime}_\mu$ 
and $\overline{b_R} \gamma^\mu s_R Z^{\prime}_\mu$ transitions. 
The presence of $\overline{\ell} \gamma^\mu \ell Z^\prime_\mu$ interactions 
from eq.~(\ref{equation:lepLag}) 
will potentially generate both $\mathcal{O}_9$ and $\mathcal{O}_9^{\prime}$ operators. 
We would like the NP contributions to $\mathcal{O}_9^{\prime}$ operator to be vanishing, which can 
be ensured  if the 2-3 element of $V^\dagger_{d_{R}} {\mathcal X}_{d_{R}} V_{d_{R}}$ vanishes.
Indeed, we would demand a stricter condition to ensure no tree-level FCNC interactions 
in the right handed $d$-type sector, i.e. $V^\dagger_{d_{R}} {\mathcal X}_{d_{R}} V_{d_{R}}$ is diagonal. 
This can be ensured if
\begin{equation}
V_{d_{R}} \approx  I \quad \text{or} \quad {\mathcal X}_{d_{R}} \propto I\;.
\label{equation:quarkconstraint}
\end{equation}        

The non-universal charge assignments in the quark sector will 
also be constrained by the observed neutral meson mixings.
In particular,  the constraints in the $K$--$\overline{K}$ oscillations are by far the most stringent, 
and severely constrain the flavour changing $Z^{\prime}$ interaction with the first two 
generation quarks.  
This can be accounted if we choose ~\cite{horizontalpaper,Allanach:2015gkd}  
\begin{equation}
X_{d_L} = X_{s_L} \;, \quad X_{d_{R}} = X_{s_{R}} \;.
\end{equation}

Another extremely important constraint stems from the requirement that the theory 
be free of any gauge anomalies. If the charge assignments are vector-like, i.e.  
\begin{eqnarray}
{\mathcal X}_{u_{L}} = {\mathcal X}_{d_{L}} = {\mathcal X}_{u_{R}} = {\mathcal X}_{d_{R}} 
\equiv {\mathcal X}_Q \quad \;,\quad
{\mathcal X}_{\ell_{L}} = {\mathcal X}_{{\nu_\ell}_{L}} 
= {\mathcal X}_{\ell_{R}} = {\mathcal X}_{{\nu_\ell}_{R} } \equiv {\mathcal X}_L  \;,
\end{eqnarray}
and are related by the condition
\be
{\rm Tr}\left[\, 3 \, {\mathcal X}_{Q} + {\mathcal X}_{L} \right] = 0\; ,
\ee
the theory is free of all gauge anomalies.
The $X$-charge assignments can then be written in a simplified notation as given
in table~\ref{table:1}. In terms of this notation, the anomaly-free condition is 
\begin{eqnarray}
3 \,(2 x_1 + x_3) + y_e + y_\mu + y_\tau = 0 \;.
\label{anomaly}
\end{eqnarray}

\begin{table}[h!]
\begin{center}
\begin{tabular}{|c|c|c|c|c|c|c|c|}
\hline
 Fields & $Q_1$ & $Q_2$  & $Q_3$ & $L_1$ & $L_2$ & $L_3$ & $\Phi$ \\ \hline
 $U(1)_X$ & $x_1$ & $x_1$ & $x_3$ & $ y_e$ & $y_\mu$ & $y_\tau$ & $0$   \\ \hline
\end{tabular}
\end{center}
\caption{\label{table:1} Vector-like $X$-charge assignments after applying preliminary 
constraints from the vanishing of NP contributions to $\mathcal{O}_9^{\prime}$, $\mathcal{O}_{10}$ and 
$\mathcal{O}_{10}^\prime$ operators, and constraints from $K$--$\overline{K}$ mixing.
Here $Q_i$ and $L_i$ represent the $i^{\rm th}$ generations of quarks and leptons, respectively.}
\end{table}

We are now in a position to select the correct alternative in eq.~(\ref{equation:quarkconstraint}).
The NP contribution to the ${\mathcal O}_9$ operator would require $x_1$ and $x_3$ to be unequal (see section~\ref{section:selection}), 
i.e. ${\mathcal X}_{d_{L}} \neq I$. The vector-like charge assignments then imply
${\mathcal X}_{d_{R}} \neq I$, and the only possibility remaining from 
eq.~(\ref{equation:quarkconstraint}) is  $V_{d_{R}} \approx I$. 
This condition need not be automatically satisfied in our model. In addition, 
${\mathcal X}_{d_{L}}={\mathcal X}_{d_{R}} \neq I$ could create problems in generating the structure of the
quark mixing matrix. We shall discuss the way to overcome these issues in section~\ref{section:ckm}.

\subsection{Enlarging the scalar sector}
\subsubsection{Additional doublet Higgs to generate the CKM matrix}
\label{section:ckm}

The Yukawa interactions of quarks with the Higgs doublet $\Phi$ are
\begin{eqnarray}
\mathcal{L}_{\rm Yuk} = \overline{Q^{\rm f}_{L}} \, {\mathcal Y}^{u} \, \Phi^c \,  u^{\rm f}_{R} 
+  \overline{Q^{\rm f}_{L}} \, {\mathcal Y}^{d} \, \Phi d^{\rm f}_{R} \;.
\label{equation:yukawa}
\end{eqnarray}
where the superscript ``f'' indicates flavour eigenstates.
The $X$-charge assignments given in table~\ref{table:1}
govern the structure of the Yukawa matrices (${\mathcal Y}^u$ and ${\mathcal Y}^d$) as 
\begin{equation}
{\mathcal Y}^{u} = \begin{pmatrix} 
\times & \times  & 0\\
\times & \times & 0\\
0 & 0 & \times 
\end{pmatrix}, \quad  
{\mathcal Y}^{d} =  \begin{pmatrix} 
\times & \times  & 0\\
\times & \times & 0\\
0 & 0 & \times 
\end{pmatrix} \; ,
\label{equation:yukawamatrices}
\end{equation}
where $\times$ denote nonzero values.
Quarks masses are obtained by diagonalizing the above ${\mathcal Y}^u$ and ${\mathcal Y}^d$ 
matrices using the bi-unitary transformations $V_{u_L}^\dagger {\mathcal Y}^u V_{u_R}$ and 
$V_{d_{L}}^\dagger {\mathcal Y}^d V_{d_{R}}$, respectively. 
Clearly, the rotations would be only in 1-2 sector. 
Therefore the quark mixing matrix, i.e. $V_{\text{CKM}} = V_{u_{L}}^\dagger V_{d_{L}}$ 
also would have non-trivial rotations only in the $1$-$2$ sector,
however this cannot be a complete picture as we know that all the elements 
of $V_{\rm CKM}$ are non-zero.

The correct form of $V_{\text{CKM}}$ can be obtained if mixings between 1-3 
and 2-3 generations are generated. This can be achieved by enlarging the scalar 
sector of SM through an addition of one more SM-like doublet, $\Phi_1$, 
with $X$-charge equal to $\pm d\,$ where $d=(x_{1} - x_{{3}})$.  
We choose $X_{\Phi_1} = + d$, similar to that in \cite{horizontalpaper}.

We first show how the 1-3 and 2-3 mixings are generated with the addition of this 
new Higgs doublet.
The generic representations for these doublets $\Phi_1$ and $\Phi_2 \equiv \Phi$ are
\begin{eqnarray*} 
\Phi_1 & = &  \begin{pmatrix}
\phi_1^+ \\
\frac{1}{\sqrt{2}} [\text{Re}(\phi_1) + i \text{Im}(\phi_1) + v_1]
\end{pmatrix}, \quad 
\Phi_2 = \begin{pmatrix}
\phi_2^+ \\
\frac{1}{\sqrt{2}} [\text{Re}(\phi_2) + i \text{Im}(\phi_2) + v_2]
\end{pmatrix},
\end{eqnarray*}
where $v_1$ and $v_2$ are vacuum expectation values of the two doublets. 
There are related by $v_1 = v \cos\beta$ and $v_2 = v \sin \beta$, 
where $v$ is the electroweak vacuum expectation value.
With this addition the Lagrangian in eq.~(\ref{equation:yukawa})
gets modified to
\begin{eqnarray}
\mathcal{L}_{\rm Yuk} =  \overline{Q^{\rm f}_{L}} \, \bigg( {\mathcal Y}_1^{u} \Phi_1^{c}
+ {\mathcal Y}^u \Phi_2^{c} \bigg)  u^{\rm f}_{R} 
+ \overline{Q^{\rm f}_{L}} \,  \bigg( {\mathcal Y}_1^{d} \Phi_1 + {\mathcal Y}^d \Phi_2 \bigg) d^{\rm f}_{R} \;,
\label{eq:Lyuk2}
\end{eqnarray}
where
\begin{equation}
{\mathcal Y}_1^{u} =   \begin{pmatrix}
0 & 0 & 0\\
0 & 0 & 0\\
\times & \times & 0\\
\end{pmatrix} , \quad  
{\mathcal Y}_1^{d} =   \begin{pmatrix}
0 & 0 & \times \\
0 & 0 & \times \\
0 & 0 & 0\\
\end{pmatrix} \;.
\label{eq:Y1-matrices}
\end{equation} 
The bi-unitary transformations would now diagonalize the quark mass matrices as
\begin{eqnarray}
M^{\text{diag}}_u &=& \frac{v}{\sqrt{2}} V^\dagger_{u_{L}} \left({\mathcal Y}^u_{1}  \cos\beta +  
{\mathcal Y}^u  \sin\beta \right) V_{u_{R}}\;, 
\label{eqn:updiagmass}  \\
M^{\text{diag}}_d &=& \frac{v}{\sqrt{2}} V^\dagger_{d_{L}} \left({\mathcal Y}^d_{1}  \cos\beta +  
{\mathcal Y}^d  \sin\beta \right) V_{d_{R}} \; . 
\label{eqn:downdiagmass}
\end{eqnarray}
From eqs.~(\ref{equation:yukawamatrices}), (\ref{eq:Lyuk2}) and (\ref{eq:Y1-matrices}),
it may be seen that rotations in 1-2, 1-3 as well as 2-3 sector will now be needed to 
diagonalize the Yukawa matrices.
Appropriate choice of parameters can then reproduce the correct form of $V_{\text{CKM}}$.
We choose $V_{u_L} = I$, so that $V_{d_L} = V_{\text{CKM}}$, which ensures that $Z^\prime$ 
does not introduce any new source of CP violation in $B$--$\overline{B}$ mixing.

Having fixed $V_{u_L}$ and $V_{d_L}$, we now turn to $V_{u_R}$ and $V_{d_R}$. 
The solution to eq.~(\ref{eqn:updiagmass})  
yields $[\mathcal{Y}_1^u]_{ij}=0$, implying the mixing angle between the 2-3 and 1-3 
generation for up type quark is zero.
The solution does not constrain the rotation angle between the first and the second generation,
which we choose to be vanishing for simplicity. Hence, $V_{u_R}$ in our model is $I$.

Note that 
eq.~(\ref{equation:quarkconstraint}) and subsequent discussion near the end of 
section~\ref{sec:prelim} led to the requirement $V_{d_R} \approx I$. 
We shall now see that this requirement is easily satisfied in this framework.
With $V_{d_L} = V_{\rm CKM}$, eq.~(\ref{eqn:downdiagmass}) may be written in the form 
\be
V_{\rm CKM} M^{\rm diag}_d V_{d_R}^\dagger = 
\begin{pmatrix}
\times & \times & \times \\
\times & \times & \times \\
0 & 0 & \times \\
\end{pmatrix} \;.
\ee
It may be seen that $V_{d_R}$ with small rotation angles, parametrized as
\begin{equation} 
 V_{d_{R}} \approx \begin{pmatrix}
 1 && \theta_{d_{R12}} && \theta_{d_{R13}} e^{-i \,\delta_d} \\
 -\theta_{d_{R12}} && 1 && \theta_{d_{R23}} \\
 -\theta_{d_{R13}}e^{i\, \delta_d} && -\theta_{d_{R23}} && 1
 \end{pmatrix} \; , 
\end{equation} 
can lead to the above form, with 
\begin{equation}
\theta_{d_{R{23}}} \approx A \lambda^2 m_s/m_b \;, \quad \theta_{d_{R{13}}} \approx - A \lambda^3 m_d/m_b \;,
\label{eq:vdr}
\end{equation}
where $A$ and $\lambda$ are the Wolfenstein parameters and $m_d$, $m_s$ and $m_b$ are the quark masses.
Note that similar observation has been made in~\cite{horizontalpaper}. 
The value of $\theta_{d_{R12}}$ is not constrained, and can be chosen to be vanishing.
Thus, the requirement $V_{d_R}\approx 1$ is satisfied.
 
Note that since $V_{d_{R}}$ is only approximately equal to $I$, small NP contributions to $C_9^\prime$ 
are present, However as we shall see in section~\ref{section:selection}, these contributions are roughly 
$(A \lambda^2 m_s)/(m_b V_{tb} V_{ts}^*)$ times the NP contributions to $C_9$, 
and hence can be safely neglected.

\subsubsection{Singlet scalar for generating neutrino masses and mixing pattern}
 \label{section:neutrino}

Our model has three right handed neutrinos, $\nu_R$'s. The Dirac and the Majorana mass terms for
neutrinos are
\begin{equation}
\mathcal{L}^{\text{mass}}_{\nu} = -\overline{\nu_L} m_D \nu_R -\frac{1}{2} \overline{\nu^c_R} M_R \nu_R + h.c. \;,
\label{eq:L-mass-nu}
\end{equation}
where the basis chosen for $\nu_L$ is such that the charged lepton mass matrix is diagonal. 
The active neutrinos would then get their masses through the Type-I seesaw mechanism. The net mass matrix 
being
\be
M_\nu = - m_D \, M_R ^{-1} \, m_D^T  \; .
\ee 
Since the neutrinos are charged under $U(1)_X$ with charges ($y_e$,$y_\mu$ and $y_\tau$),
 the $(\alpha,\beta)$ elements 
of the Dirac mass matrix $m_D$ would be nonzero only when $y_\alpha = y_\beta$, 
while the $(\alpha,\beta)$ elements of the
Majorana mass matrix $M_R$ would be nonzero only when $y_\alpha + y_\beta =0$. 
Since the $X$-charges of neutrinos are non-universal and vector-like, the former condition 
implies that $m_D$ is diagonal. (It can have off-diagonal elements if two of the $y_\alpha$'s 
are identical. However we can always choose the $\nu_R$ basis such that $m_D$ is diagonal.) 
The allowed elements of $M_R$ are also severely restricted, and it will not be possible to have 
a sufficient number of nonzero elements in $M_R$ to be able to generate the neutrino mixing pattern. 

To generate the required mixing pattern, we introduce a scalar $S$, which is a SM-singlet, 
and has an $X$-charge $X_S =a$, as a minimal extension of our model.
With the addition of this scalar, the Lagrangian in eq.~(\ref{eq:L-mass-nu}) modifies to   
\begin{eqnarray}
[\mathcal{L}^{\text{mass},S}_{\nu}]_{\alpha\beta} &=& [\mathcal{L}^{\text{mass}}_{\nu}]_{\alpha\beta}
             -\frac{1}{2} [\overline{\nu^c_{R}}]_\alpha [{\mathcal Y}_{R}]_{\alpha\beta} [\nu_{R}]_\beta \,S + h.c. \;.
\end{eqnarray}
The conditions for $m_D$, $M_R$ and ${\mathcal Y}_R$ elements to be non zero are
\begin{eqnarray} 
{[m_D]}_{\alpha\beta} &\neq & 0 \quad \text{if} \quad y_\alpha - y_\beta= 0 \;,  \\
\label{equation:dirac}
{[M_R]}_{\alpha\beta} &\neq& 0  \quad \text{if} \quad y_\alpha + y_\beta = 0 \;, \nonumber \\
{[{\mathcal Y}_R]}_{\alpha\beta} &\neq& 0  \quad \text{if} \quad y_\alpha + y_\beta = \pm a \;.
\label{equation:majorana}
\end{eqnarray}
When $S$ gets a vacuum expectation value $v_{S}$, it contributes to the Majorana mass term for right handed 
neutrinos which now becomes
 \begin{equation}
  {[M^{S}_R]}_{\alpha\beta} = {[M_R]}_{\alpha\beta} + \frac{v_S}{\sqrt{2}}[y_{R}]_{\alpha\beta} \;.
  \label{eq:seesaw}
 \end{equation}
Thus an element of ${[M^{S}_R]}_{\alpha\beta}$ will be non-zero if,
\begin{equation}
y_\alpha + y_\beta = 0, \pm a \;.
\label{equation:MRs}
\end{equation} 

The textures in the neutrino mass matrix, i.e. the number and location of vanishing elements
therein, hold clues to the internal flavour symmetries.
Only some specific textures of $M_R$ are allowed. While no three-zero textures are consistent
with data, specific two-zero textures are allowed~\cite{Lavoura:2004tu,2minors,textures,Heeck:2013bla}. In addition, 
most one-zero textures~\cite{1minors}, and naturally, all no-zero textures, are also permitted.
Among the allowed textures, we identify those that can be generated by a $U(1)_X$ symmetry
with a singlet scalar, i.e. those for which values of $y_\alpha$ and $a$ satisfying 
eq.~(\ref{equation:MRs}) may be found. These combinations are listed in table~\ref{table:texture1}, 
and categorized according to the ratio $y_e/y_\mu$. Note that by the leptonic symmetry 
combination $p_e L_e + p_\mu L_\mu + p_\tau L_\tau$, we refer to all $U(1)_X$ charge combinations,
where $p_e/y_e = p_\mu/ y_\mu = p_\tau/ y_\tau$ (for non zero values $y_\alpha$ and $p_\alpha$ respectively). 
It is to be noted that  part of the list was already derived in \cite{textures,Heeck:2013bla}. 
Later in section~\ref{section:selection}, we shall examine the consistency of these symmetries
with the flavour data.

Note that we would like all the elements of right handed neutrino mass matrix to have 
 similar magnitudes, so it would be natural to have $[M_R^S]_{\alpha, \beta}\sim \mathcal{O}({v_S})$.
Our scenario is thus close to a TeV-scale seesaw mechanism~\cite{Atwood:2007zza}.
\begin{table}[t]
\begin{center}
\def\arraystretch{1.5}
\begin{tabular}{|c|c|c|c|c|} 
\hline
Category & $y_e$/a & $y_\mu/a$ & $y_\tau/a$ & Symmetries\\ \hline
A & $0$ & $-1$ & $0$, $1$ & $L_\mu$, $L_\mu-L_\tau$\\ \hline
B & $\frac{1}{2}$ & $-\frac{3}{2}$ & $\pm\frac{1}{2}$  & $L_e - 3 L_\mu \pm L_\tau$\\ \hline
C & $-\frac{1}{2}$ & $-\frac{3}{2}$ & $\frac{1}{2}$  & $L_e + 3 L_\mu - L_\tau$\\ \hline
D & $\frac{1}{2}$ & $-\frac{1}{2}$ &  $\pm\frac{1}{2}$, $\pm\frac{3}{2}$  
& $L_e -L_\mu \pm L_\tau$, $L_e - L_\mu \pm 3 L_\tau$\\ \hline
E & $\frac{1}{2}$ & $ \frac{1}{2}$ & $-\frac{1}{2}$, $-\frac{3}{2}$  
&$L_e + L_\mu - L_\tau$, $L_e + L_\mu -3 L_\tau$ \\ \hline
F & $\frac{3}{2}$ & $-\frac{1}{2}$ & $-\frac{1}{2}$  & $3 L_e - L_\mu - L_\tau$\\ \hline
G & $1$ & $0$ & $0$  & $\,L_e$\\ \hline
\end{tabular}
\caption{\label{table:texture1} 
The $X$-charges (in units of $a$) along with the symmetry combinations that are consistent with 
the neutrino oscillation data~\cite{Capozzi:2016rtj, Esteban:2016qun}. Note that by the leptonic symmetry 
combination $p_e L_e + p_\mu L_\mu + p_\tau L_\tau$, we refer to all $U(1)_X$ charge combinations,
where $p_e/y_e = p_\mu/ y_\mu = p_\tau/ y_\tau$ (for non zero values $y_\alpha$ and $p_\alpha$ respectively). 
In the list we have dropped the cases with lepton flavour universality and the one where 
$y_e=y_\mu=0$.}
\end{center}
\end{table}

\subsubsection{Relating $X$-charges of doublet and singlet scalars}  
\label{sec:scalars}

The scalar sector of our model consists of two $SU(2)_L$ doublets $\Phi_1$ and $\Phi \equiv \Phi_2$, and a 
SM-singlet $S$, with $X$-charges $d$, 0, $a$, respectively.
The scalar potential that respects the $SU(2)_L \times U(1)_Y \times U(1)_X$ symmetry is
\begin{eqnarray}
V_{\Phi_1\Phi_2S} &=&  - m_{11}^2 \Phi_1^\dagger\Phi_1 + \frac{\lambda_1}{2} (\Phi_1^\dagger \Phi_1)^2 
- m_{22}^2 \Phi_2^\dagger\Phi_2 + \frac{\lambda_2}{2} (\Phi_2^\dagger \Phi_2)^2 \nonumber \\ 
&&  -m_{S}^2 S^\dagger S   + \frac{\lambda_{S}}{2} (S^\dagger S)^2
 + \lambda_3 \, \Phi_1^\dagger \Phi_1 \, \Phi_2^\dagger \Phi_2  \nonumber \\
 &&  + \lambda_4 \,\Phi_1^\dagger \Phi_2 \, \Phi_2^\dagger \Phi_1 
 + \left(\lambda_{1S{_1}} \Phi_1^\dagger \Phi_1 + \lambda_{2{S}}  \Phi_2^\dagger \Phi_2\right) S^\dagger S\;.
\end{eqnarray}
The $U(1)_X$ symmetry is broken spontaneously by the vacuum expectation values of $\Phi_1$ and $S$, 
and consequently $Z^\prime$ obtains a mass (see appendix~\ref{section:zprime}). 
Since the collider bounds indicate $M_{Z^\prime} \gtrsim$ TeV, we expect $v_s \gtrsim$ TeV 
(since $v_1 \lesssim$ electroweak scale).  

Therefore, before electroweak symmetry breaking, $U(1)_X$ symmetry gets broken spontaneously and the singlet, $S$,
gets decoupled. The effective potential for the doublets after $U(1)_X$ symmetry breaking
\begin{eqnarray}
  V_{\Phi_1\Phi_2} &=&    - \left(m_{11}^2 -\frac{\lambda_{1S{_1}}}{2} v^2_{S}  \right) \Phi_1^\dagger\Phi_1 
           + \frac{\lambda_1}{2} (\Phi_1^\dagger \Phi_1)^2 
           - \left(m_{22}^2 - \frac{\lambda_{2{S}}}{2} v^2_{S}\right) \Phi_2^\dagger\Phi_2 \nonumber \\
          && + \frac{\lambda_2}{2} (\Phi_2^\dagger \Phi_2)^2 
             + \lambda_3 (\Phi_1^\dagger \Phi_1) (\Phi_2^\dagger \Phi_2) + \lambda_4 (\Phi_1^\dagger \Phi_2) (\Phi_2^\dagger \Phi_1) .
\end{eqnarray}
The potential, $V_{\Phi_1,\Phi_2}$, is invariant under the global transformation
$U(1)_V \times U(1)_A$ such that
\begin{eqnarray}
U(1)_V \times U(1)_A : \quad \Phi_1 &\rightarrow& e^{i (\theta_V-\theta_A)} \Phi_1, \quad \quad
\Phi_2 \rightarrow e^{i (\theta_V + \theta_A)} \Phi_2.
\end{eqnarray}
Out of $U(1)_V$ and $U(1)_A$, only $U(1)_V$ can be gauged and identified as $U(1)_Y$ since both 
the doublets should have the same hypercharge. After electro-weak symmetry breaking, along 
with the gauge symmetries, $U(1)_A$ would also be broken spontaneously and would result in 
a Goldstone boson. This problem would not arise if the potential were not symmetric under 
$U(1)_A$ to begin with, i.e. if it were broken explicitly by a term
\begin{equation}
\Delta V_{\Phi_1\Phi_2} = - m_{12}^2\Phi_1^\dagger \Phi_2 \, + h.c. \;.
\end{equation} 
Note that this can happen naturally in our scenario: the term above can be generated by 
spontaneously breaking of $U(1)_X$ if $X_S$ is equal to $X_{\Phi_1}$, i.e., if $a=d$, 
we can have 
\begin{eqnarray}
\Delta V_{\Phi_1\Phi_2S} &=&  -\widetilde{m}_{{12}} \left[ S \, \Phi_1^\dagger \Phi_2 
+ S^\dagger \, \Phi_2^\dagger \Phi_1\right] \; ,
\label{phi1phi2}
\end{eqnarray}
with
\begin{equation}
m_{12}^2 = \frac{1}{\sqrt{2}} \widetilde{m}_{{12}} v_{S} \;.
\end{equation}
Thus the identification $X_S = X_{\Phi_1} = a$ naturally avoids a massless scalar in our model
by modifying the potential as 
 \begin{equation}
V_{\Phi_1\Phi_2S} \to V_{\Phi_1\Phi_2S} + \Delta V_{\Phi_1\Phi_2S}\;.
\label{equation:Vpot}
\end{equation}

\subsection{Selection of the desirable symmetry combinations}
\label{section:selection}

\begin{figure} [t]
\begin{center}
\includegraphics[totalheight=8cm]{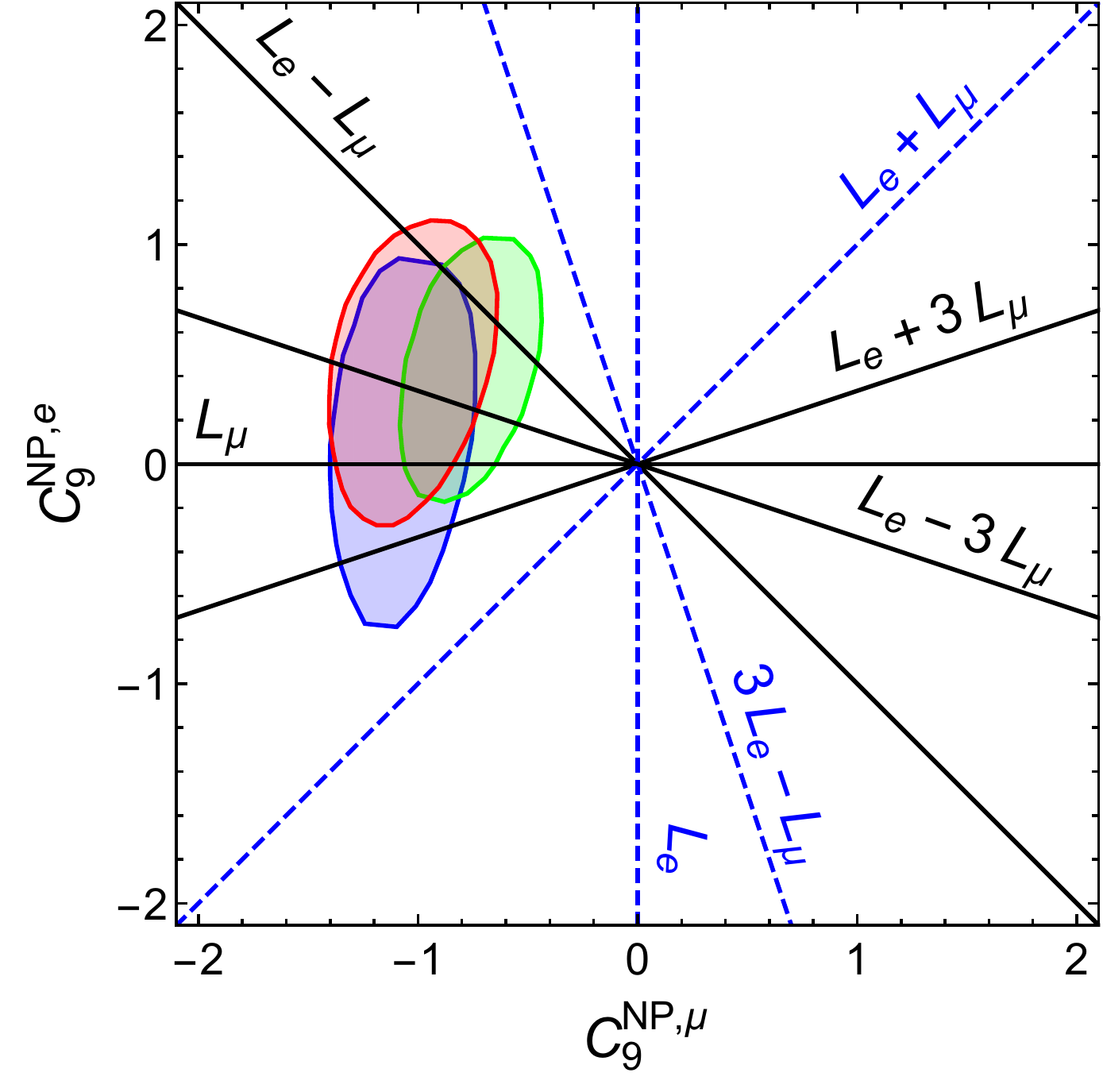}
\caption{Allowed $1\sigma$ regions in ($C_9^{\text{NP},e}$,$C_{9}^{\text{NP},\mu}$) plane 
using the global fit data: red contour is obtained from~\cite{globalfit2}, blue from~\cite{globalfit3}
and green from~\cite{globalfit4}. Lines for various $U(1)_X$ symmetries using eq.~(\ref{eq:relationbtC9}) 
have also been plotted. We do not show $\tau$ charge explicitly in the plot. }
       \label{fig:combinedfits}
       \end{center}
       \end{figure}

In this section, we combine the $U(1)_X$ symmetries  identified in section~\ref{section:neutrino}
with the NP contribution to ${\mathcal O}_9$ needed to account for the flavour anomalies.
The Lagrangian describing the $Z^\prime$ interactions with $d$-type quarks and 
charged leptons is
\begin{eqnarray}
\mathcal{L}_{Z^\prime} & = & g_{Z^\prime} \,\overline{d_L}  \gamma^\mu \, V^\dagger_{\text{CKM}} \,
{\mathcal X}_{Q} \, V_{\text{CKM}} d_L\,  Z^\prime_\mu
+ g_{Z^\prime} \, \overline{d_R}\, \gamma^\mu \, V^\dagger_{d_R}  \, {\mathcal X}_{Q} \,
V_{d_R}\, d_R \,Z^\prime_\mu \nonumber \\ 
& &  + g_{Z^\prime} \,\overline{\ell} \, \gamma^\mu \,  {\mathcal X}_{L} \,  \ell\, Z^\prime_\mu 
\end{eqnarray}
Here ${\mathcal X}_{Q}=\text{diag}(x_1,x_1,x_3)$ and ${\mathcal X}_{L}=\text{diag}(y_e,y_\mu,y_\tau)$.
Using the above Lagrangian, the $Z^\prime$ contributions to the effective Hamiltonian 
for $b \to s \ell \ell$ processes at $M_{Z^\prime}$ scale is
\begin{equation}
\mathcal{H_\text{eff}^{\text{NP}}} = -\frac{(x_1-x_3)\, y_\ell \, g^2_{Z^\prime}}{M^2_{Z^\prime}} V_{tb} V_{ts}^*
\left( \overline{s_L}  \gamma^\mu b_L \right) \left( \overline{\ell}  \gamma_\mu  \ell \right) 
+ \frac{(x_1-x_3) \, y_\ell \, g^2_{Z^\prime}}{M^2_{Z^\prime}} \theta^2_{d_{R23}}
\left( \overline{s_R}  \gamma^\mu b_R \right) \left( \overline{\ell}  \gamma_\mu  \ell \right) \; .
    \end{equation}
Comparing it with the standard definition of $\mathcal{H_{\text{eff}}}$ as given in eq.~(\ref{heff}), 
we obtain the NP contribution to the Wilson coefficients 
$C_9^{{\rm NP},\ell}$ and $C_9^{\prime \, {\rm NP},\ell}$ as 
\begin{equation}
 C_9^{\text{NP},\ell}(M_Z{^\prime}) =  \frac{\sqrt{2} \pi (x_1-x_3) \,y_{\ell} g^2_{Z^\prime}}{G_F M^2_{Z^\prime} \alpha_e }\;, 
\quad C_9^{\prime \, {\rm NP}, \ell}(M_Z{^\prime}) = -  
\frac{\sqrt{2} \pi (x_1-x_3) \,y_{\ell} g^2_{Z^\prime} \theta^2_{d_{R23}} }{G_F M^2_{Z^\prime} \alpha_e V_{tb} V_{ts}^*}\;, 
\label{equation:globalfit}
 \end{equation}
The smallness of $\theta_{R23}$, as shown in eq~(\ref{eq:vdr}), makes the NP contribution
to ${\mathcal O}_9^\prime$ small in comparison to the corresponding contribution to ${\mathcal O}_9$:
\begin{eqnarray}
 C_9^{\prime \ell}(M_Z{^\prime}) &=&  - \frac{\theta^2_{D_{R23}}}{V_{tb} V_{ts}^*} \,\,C_9^{\text{NP},\ell}(M_Z{^\prime})
 \nonumber \\
 &&\approx -0.025 \,C_9^{\text{NP},\ell}(M_Z{^\prime}) \; .
\label{equation:C9prime}
 \end{eqnarray}
The flavour anomalies like $R_K$ and $P_5^\prime$ depend crucially on $C_9^{\text{NP},e}$ and 
$C_9^{\text{NP},\mu}$, and not on $C_9^{\text{NP},\tau}$. 
A negative value of $C_9^{\text{NP},\mu}$ is preferred~\cite{Descotes-Genon:2013wba,globalfit1,globalfit2,globalfit3,globalfit4} 
as a solution to these anomalies which can be easily obtained if, $ (x_1-x_3)\, y_\mu<0 $.
The values of $C_9^{\text{NP},e}$ and $C_9^{\text{NP},\mu}$ are related by  
\begin{equation}
C_9^{\text{NP},e} / C_9^{\text{NP},\mu} = y_e / y_\mu \;.
\label{eq:relationbtC9}
\end{equation}
This ratio stays the same at all scales between $M_{Z^\prime}$ and $m_b$, since 
the $\mathcal{O}_9$ operator does not mix with any other operator at one loop in QCD.
This ratio is represented in figure~\ref{fig:combinedfits} by lines corresponding to 
different symmetries in table~\ref{table:texture1}. 

In figure~\ref{fig:combinedfits}, we also show the 1$\sigma$ contours in the $C_9^{\text{NP},\mu}$--$C_9^{\text{NP},e}$ 
plane obtained from the global fits~\cite{globalfit2,globalfit3,globalfit4}. 
For further analysis, we select only those combinations (categories A, B, C, D) 
which pass through the 1$\sigma$ regions of any of these global fit contours. 
Among these possibilities, $L_\mu -L_\tau$ has already been considered in the context of 
$R_K$ \cite{Crivellin:2015mga,horizontalpaper,Fuyuto:2015gmk,Altmannshofer:2016jzy}, where the NP contribution to $C_9^e$ is absent. 
We shall explore the phenomenological consequences of these symmetries in 
section~\ref{section:exp_constraints}.

\begin{table}[t]
\def\arraystretch{1.5}
\begin{center}
\begin{tabular}{ |c|c|c|c|c|c|c|c|} 
 \hline
 Category & Symmetry/Charges & $x_1/a$ & $x_2/a$  & $x_3/a$ & $y_e/a$ & $y_\mu/a$ & $y_\tau/a$ \\ \hline 
 A &$ L_\mu- L_\tau$ & $\frac{1}{3}$ & $\frac{1}{3}$ & $-\frac{2}{3}$ & $ 0$ & $-1$ & $1$  \\
  & $L_\mu$ & $\frac{4}{9}$ & $\frac{4}{9}$ & $-\frac{5}{9}$ & $0$ & $-1$ & $0$  \\ \hline 
 B &$L_e-3L_\mu + L_\tau$ & $\frac{7}{18}$ & $\frac{7}{18}$ & $-\frac{11}{18}$ & $ \frac{1}{2}$ & $-\frac{3}{2}$ & $\frac{1}{2}$  \\ 
 &$L_e-3L_\mu - L_\tau$ & $\frac{1}{2}$ & $\frac{1}{2}$ & $-\frac{1}{2}$ & $ \frac{1}{2}$ & $-\frac{3}{2}$ & $-\frac{1}{2}$  \\ \hline
C & $L_e + 3 L_\mu - L_\tau$ & $\frac{1}{2}$ & $\frac{1}{2}$ & $-\frac{1}{2}$ & $ -\frac{1}{2}$ & $-\frac{3}{2}$ & $\frac{1}{2}$  \\ \hline 
 D & $L_e-L_\mu +3 L_\tau$ & $\frac{1}{6}$ & $\frac{1}{6}$ & $-\frac{5}{6}$ & $ \frac{1}{2}$ & $-\frac{1}{2}$ & $\frac{3}{2}$  \\ 
  & $L_e-L_\mu -3 L_\tau$ & $\frac{1}{2}$ & $\frac{1}{2}$ & $-\frac{1}{2}$ & $ \frac{1}{2}$ & $-\frac{1}{2}$ & $-\frac{3}{2}$  \\ 
  &$L_e-L_\mu + L_\tau$ & $\frac{5}{18}$ & $\frac{5}{18}$ & $-\frac{13}{18}$ & $ \frac{1}{2}$ & $-\frac{1}{2}$ & $\frac{1}{2}$  \\ 
 & $L_e-L_\mu - L_\tau$ & $\frac{7}{18}$ & $\frac{7}{18}$ & $-\frac{11}{18}$ & $ \frac{1}{2}$ & $-\frac{1}{2}$ & $-\frac{1}{2}$  \\ \hline
 \end{tabular}
\end{center}
\caption{\label{table:selected} 
Charges of the fermion fields in units of $a$. It can be seen that for all the allowed symmetries we have $(x_1-x_3)\,y_\mu<0$. } 
\end{table}

Note that although we refer to the symmetries by their lepton combinations, quarks are also charged
under the $U(1)_X$. These charges can be easily obtained from the anomaly eq.~(\ref{anomaly}), 
and have been given in table~\ref{table:selected}, in terms of the parameter $a$.
Further, note that all the $X$-charges are proportional to $a$. As a result, $a$ and $g_{Z^\prime}$
always appear in the combination $a g_{Z^\prime}$.
We therefore absorb $a$ in the definition of $g_{Z^\prime}$: 
\begin{equation}
g_{Z^\prime} \to a \, g_{Z^\prime}\;,
\label{eqn:a}
\end{equation} 
and consider $a=1$ without loss of generality for our further analysis. 
The interactions of $Z^\prime$ then can be expressed in terms of two unknown parameters, 
$g_{Z^\prime}$ and $M_{Z^\prime}$. In the next section, we shall subject all the symmetry combinations
in table~\ref{table:selected} to tests from experimental constraints.

\section{Experimental Constraints}
\label{section:exp_constraints}

Our class of models will be constrained from flavour data and direct searches at the colliders. 
We choose to work in the decoupling regime where the 
additional scalars are heavy and do not play any significant role in the phenomenology. 
This is easily possible by suitable choice of the parameters in eq.~(\ref{equation:Vpot}). 
This framework naturally induces $Z-Z^\prime$ mixing at tree level, which can also be minimized 
by the choice of these parameters (appendix~\ref{section:zprime}).
The two parameters that are strongly constrained from the data are the mass and gauge coupling 
of the new vector boson, $Z^\prime$. In this section, we explore the constraints on 
$M_{Z^\prime}$ and $g_{Z^\prime}$ from neutral meson mixings, rare $B$ decays, and 
direct $Z^\prime$ searches at colliders.

\subsection{Constraints from neutral meson mixings and rare $B$ decays}

The FCNC couplings of $Z^\prime$ to $d_L$-type quarks (note that $V_{d_L} = V_{\rm CKM}$)
will lead to neutral meson mixings as well as $b \to d$  and $b \to s$ transitions
at the tree level, and hence may be expected to
give significant BSM contributions to these processes.

The effective Hamiltonian in SM~\cite{Buras_review} that leads to 
$K-\overline{K}$, $B_d - \overline{B_d}$ and $B_s -\overline{B_s}$ mixing is
\begin{eqnarray}
\mathcal{H}^{\text{SM}}_{\text{eff}} & = & \frac{G_F^2}{16 \pi^2} M_W^2 C^{\text{SM}}_{K}(\mu) 
\big[\overline{s} \gamma^\mu (1-\gamma_5) d\big] 
\big[\overline{s} \gamma_\mu (1-\gamma_5) d\big] \nonumber \\
 & + & \frac{G_F^2}{16 \pi^2} M_W^2 \left(V_{tb}V^{*}_{td}\right)^2 C^{\text{SM}}_{B_d}(\mu) 
\left[\overline{b} \gamma^\mu (1-\gamma_5) d\right] 
\left[\overline{b} \gamma_\mu (1-\gamma_5) d\right] \nonumber \\
 & + & \frac{G_F^2}{16 \pi^2} M_W^2 \left(V_{tb}V^{*}_{ts}\right)^2 C^{\text{SM}}_{B_s}(\mu) 
 \left[\overline{b} \gamma^\mu (1-\gamma_5) s\right] 
\left[\overline{b} \gamma_\mu (1-\gamma_5) s\right]\;,
\end{eqnarray}
where $C^{\text{SM}}_P(\mu)$ are the Wilson coefficients at the scale $\mu$ for $P=K,B_d,B_s$ 
and the CKM factors for $K$--$\overline{K}$ mixing are absorbed in $C^{\text{SM}}_{K}(\mu)$
itself.  

Contributions due the $Z^\prime$ exchange will have the same operator form as in the SM since 
(i) The FCNC contributions to $\overline{d_{Ri}} \gamma^\mu  d_{Rj} Z^\prime_\mu$ operator are 
small as shown in eqs.~(\ref{eq:vdr}) and (\ref{equation:C9prime}), and 
(ii) we are working in the decoupling limit, where the contributions due to the exchanges of scalars 
$H^0$, $A^0$ and $H^+$ are negligible (see appendix \ref{section:FCNC}).
As a result, the total effective Hamiltonian can simply be written with the replacement
\be
C^{\rm SM}_P(\mu)  \to  C^{\rm tot}_P(\mu) = C^{\rm SM}_P(\mu) + C^{\rm NP}_P(\mu) \; ,
\ee 
with the Wilson coefficients $C^{\rm NP}_P$ at the $M_{Z^{\prime}}$ scale given by
\begin{eqnarray}
C^{\text{NP}}_{K} (M_{Z^{\prime}})    &=&   
\frac{{2 \pi^2 \,(x_1-x_3)^2\,g^2_{Z^\prime} \left(V_{td}V^{*}_{ts}\right)^2}}{ M^2_{Z^\prime}G_F^2 M_W^2} \;, \nonumber \\
C^{\text{NP}}_{B_q} (M_{Z^\prime})    &=& 
\frac{{2 \pi^2 \,(x_1-x_3)^2\,g^2_{Z^\prime} }}{ M^2_{Z^\prime}G_F^2 M_W^2} \quad \rm{where,}\, q = d,s \;.
\end{eqnarray}
These Wilson coefficients at one loop in QCD run down to the $M_W$ scale as~\cite{Buras_review}
\begin{equation}
C^{\text{NP}}_{P} (M_W)  = \left[\frac{\alpha_s(m_t)}{\alpha_s(M_W)}\right]^{\frac{6}{23}}\,
\left[\frac{\alpha_s(M_{Z^\prime})}{\alpha_s(m_t)}\right]^{\frac{2}{7}} 
C^{\text{NP}}_{P} (M_{Z^\prime}) \;.
\label{running}
\end{equation}
Since the form of operators corresponding to $C^{\text{NP}}_{P}(\mu)$ and $C^{\text{SM}}_P(\mu)$ 
is the same, the ratio $C^{\text{NP}}_{P}(\mu)/C^{\text{SM}}_P(\mu)$ stays the same for all scales 
below $M_W$. Since only this ratio is relevant for the constraints from $P$--$\overline{P}$
mixing, we work in terms of $C^{\text{NP}}_{P}(M_W)/C^{\text{SM}}_P(M_W)$.

\begin{figure} [t]
\begin{center} 
 \includegraphics[totalheight=6cm]{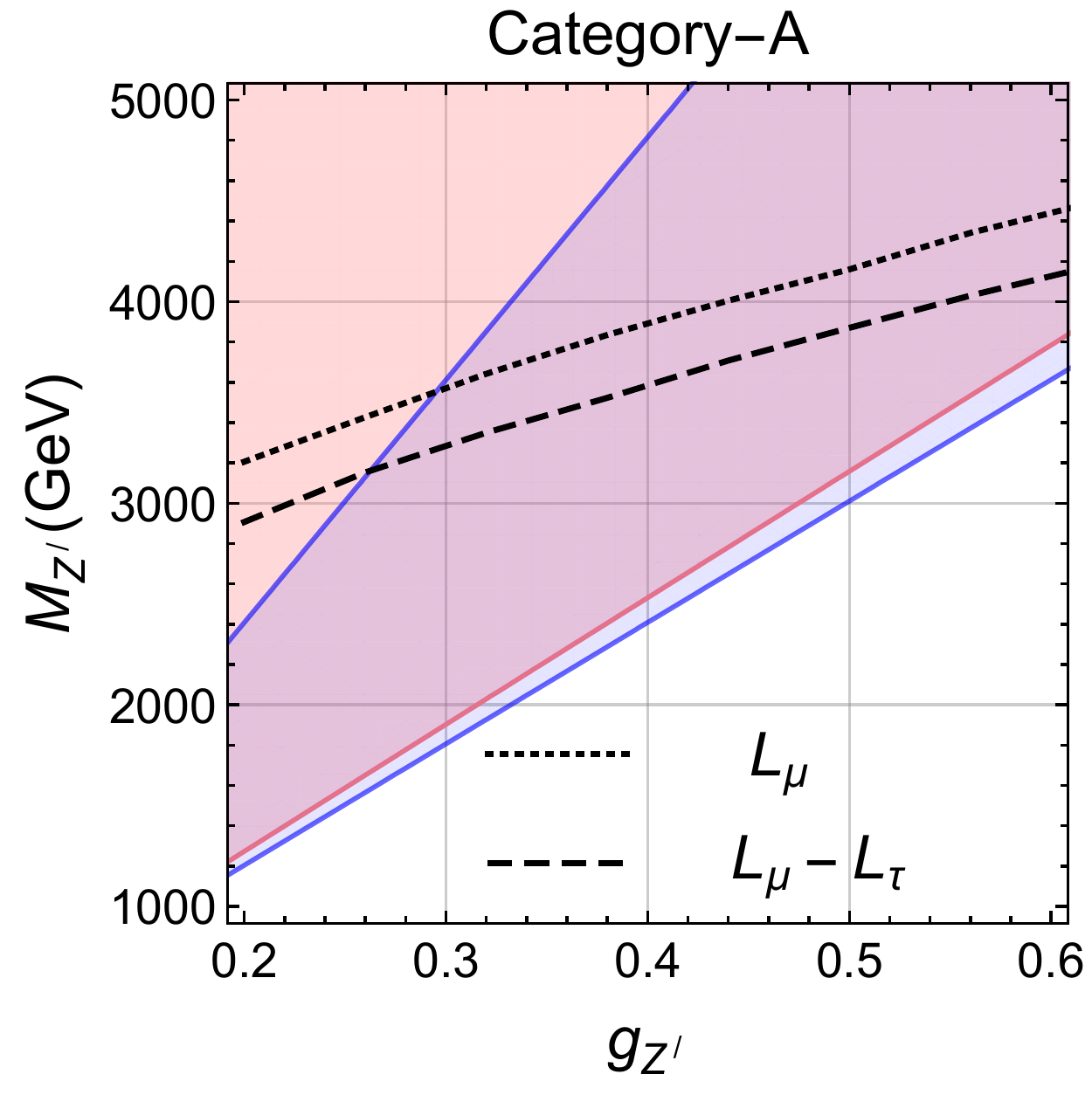}
 \quad \includegraphics[totalheight=6cm]{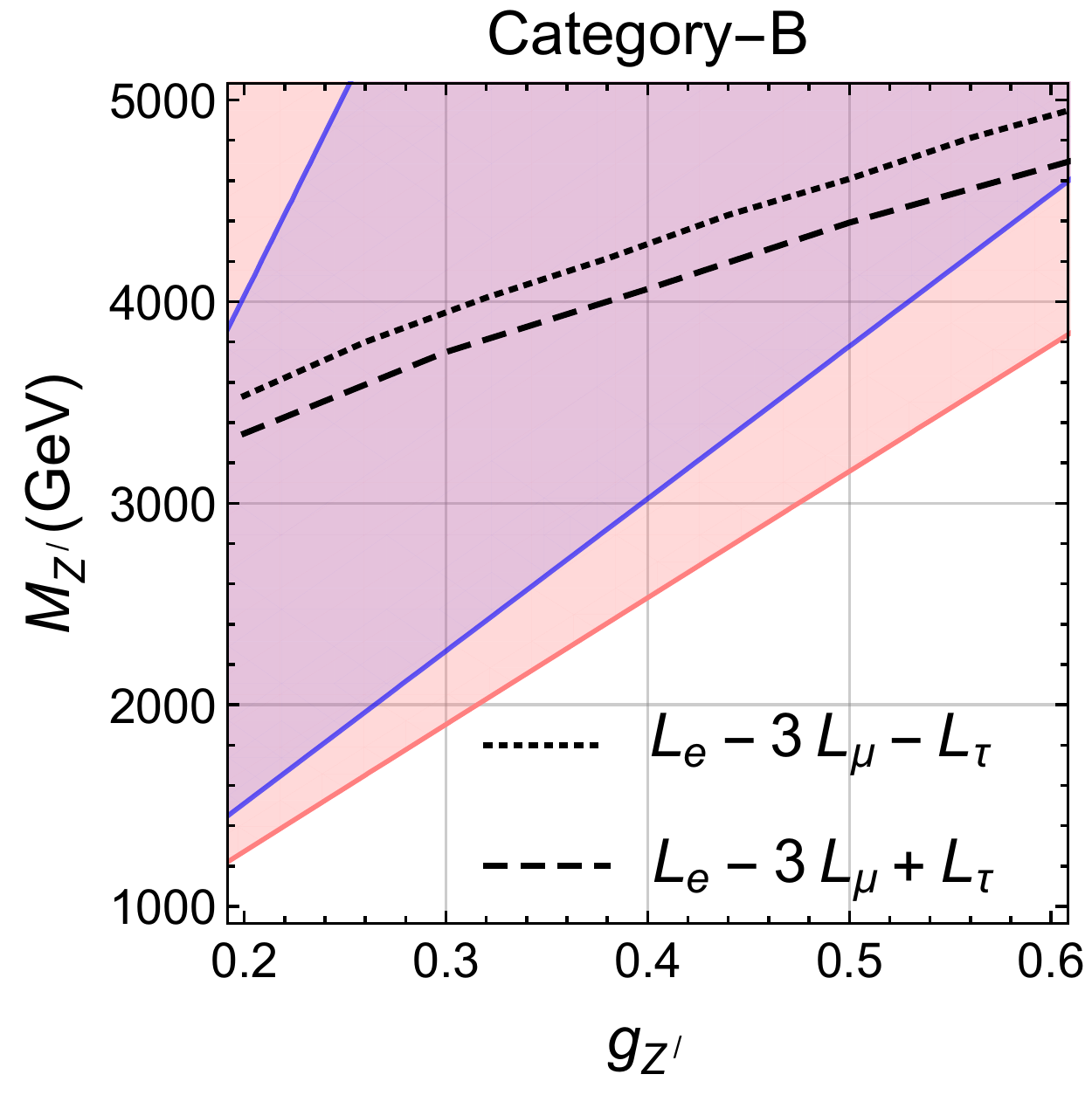}\\
 \includegraphics[totalheight=6cm]{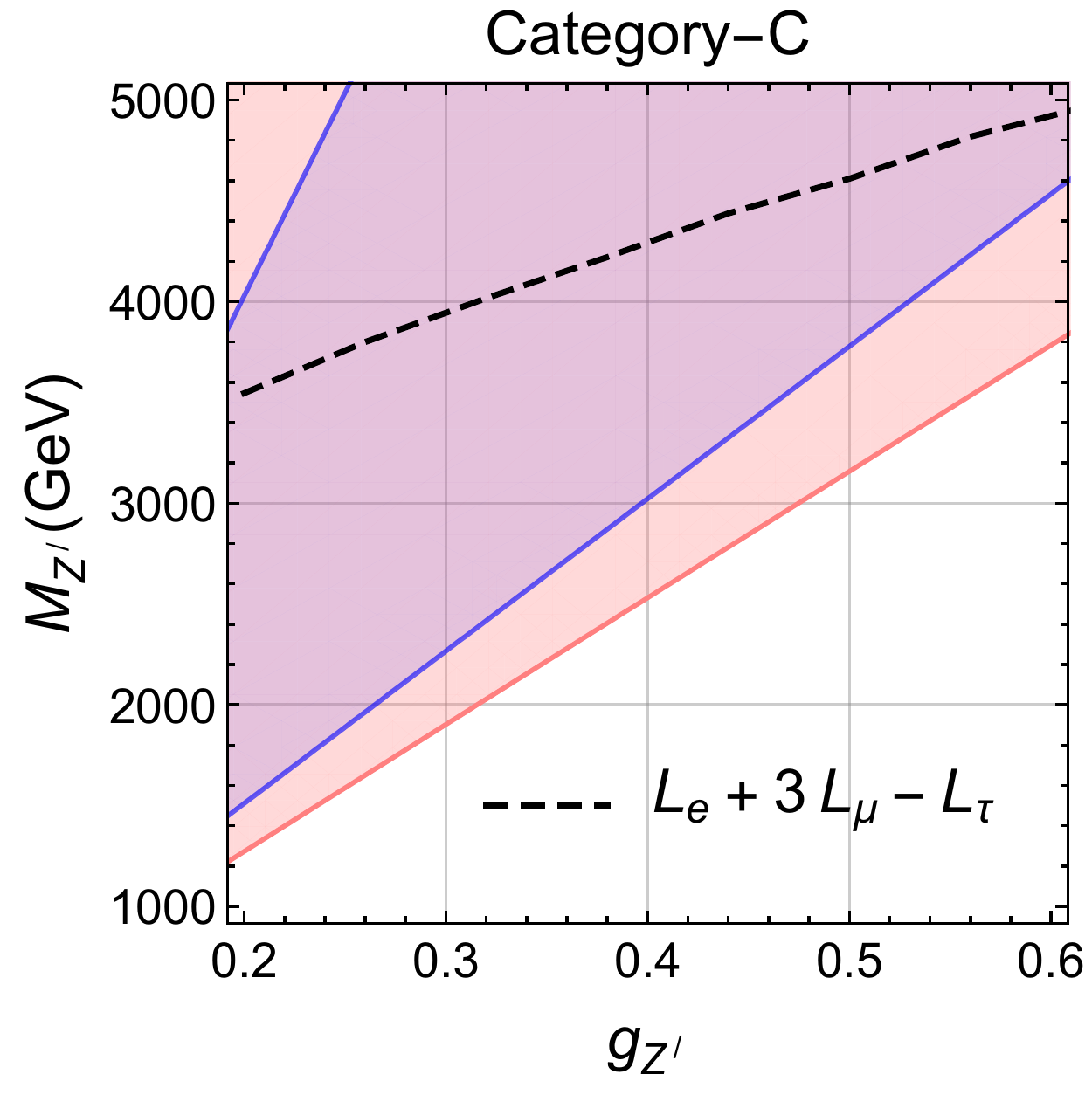}
 \quad \includegraphics[totalheight=6cm]{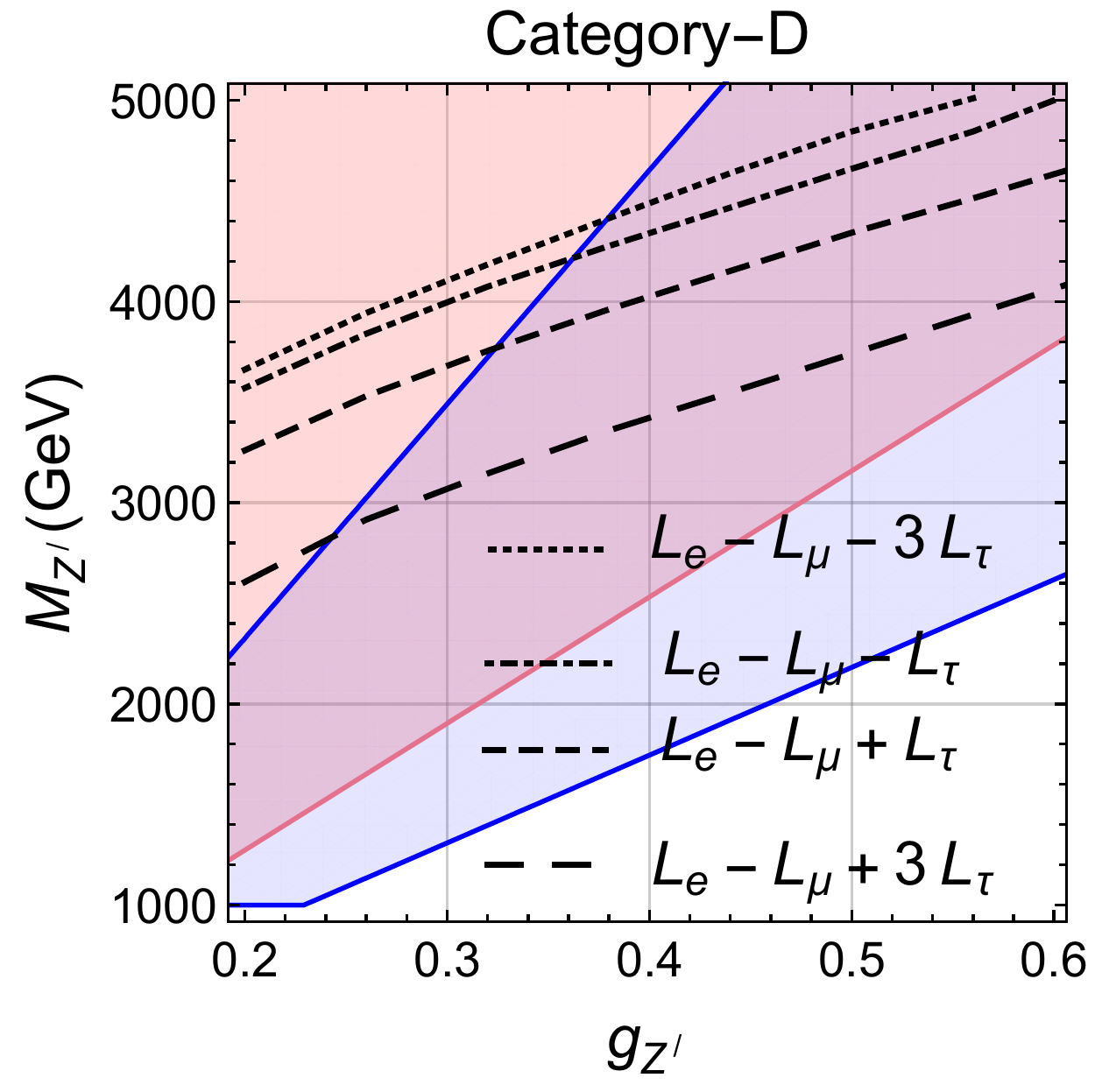}
 \caption{\label{fig:combinedconstraints}
The constraints in the $g_{Z^\prime}$ --$M_{Z^\prime}$ plane, from neutral meson mixings, 
rare $B$ decays, and collider searches for $Z^\prime$, for the symmetry categories
in table~\ref{table:selected}.
The 2$\sigma$ regions allowed by the neutral meson mixings are shaded pink, 
while the 2$\sigma$ regions allowed by the global fit \cite{globalfit4} to $b \to s\ell\ell$
and $b \to s\gamma$ is shaded blue. Purple is the overlap of these two constraints.
The dotted and dashed lines correspond to the collider bounds -- the regions above them
are allowed at 95\% C. L.. The net allowed region for a given symmetry is therefore the 
purple region lying above the dotted / dashed line corresponding to that symmetry. 
}
\end{center} 
\end{figure}

The constraints from $P$--$\overline{P}$ measurements are generally parametrized 
in terms of the following quantities~\cite{UTFit}:
\be
C_{\epsilon_K} \equiv  \frac{\text{Im}\left[\left\langle K_0| \mathcal{H}^{\text{tot}}_{\text{eff}}|  
\bar{K_0}\right\rangle\right]} {\text{Im}\left[\left\langle K_0| \mathcal{H}^{\text{SM}}_{\text{eff}}|  
\bar{K_0}\right\rangle\right]} \;, \quad
C_{B_q} e^{2i\phi_{B_{q}}} \equiv  \frac{\left\langle B_q| \mathcal{H}^{\text{tot}}_{\text{eff}}|  
\bar{B_q}\right\rangle}{\left\langle B_q| \mathcal{H}^{\text{SM}}_{\text{eff}}| \bar{B_q}\right\rangle}\;. 
\ee
Note that the quantity 
$C_{\Delta{m_K}} \equiv \text{Re}\left[\left\langle K_0| \mathcal{H}^{\text{tot}}_{\text{eff}}|  
\bar{K_0}\right\rangle\right]/\text{Re}\left[\left\langle K_0| \mathcal{H}^{\text{SM}}_{\text{eff}}|  
\bar{K_0}\right\rangle\right]$ is also a relevant observable, however since it receives large 
long distance corrections, we do not consider it 
in our analysis. Since $V_{D_{L}} = V_{\text{CKM}}$, there is no new phase contributions to 
$B_q-\overline{B_q}$ mixing and $\phi_{B_q} = 0$.

We combine the above measurements and show the allowed 2$\sigma$ regions
in the $g_{Z^\prime}$--$M_{Z^\prime}$ plane in figure~\ref{fig:combinedconstraints}.
Note that constraints from neutral meson mixings depends on $g_{Z^\prime}$, $M_{Z^\prime}$
and $(x_1-x_3)$. Since $(x_1-x_3)=a$, therefore the 
$P$--$\overline{P}$ constraints are the same in all the categories in table~\ref{table:selected}
(and hence for all the four panels of figure~\ref{fig:combinedconstraints}).

Figure~\ref{fig:combinedconstraints} also shows the 2$\sigma$ allowed regions that
correspond to the constraints from a global fit \cite{globalfit4} incorporating the 
$b \to s \ell \ell$ and $b \to s \gamma$ data. Note that these constraints
have already been used in shortlisting the lepton symmetries in table~\ref{table:selected},
Here we find the allowed regions in the $g_{Z^\prime}$--$M_{Z^\prime}$ plane
using eq.~(\ref{equation:globalfit}). The constraints depend on the $X$-charges of the electron 
and muon, but are independent of the charge of $\tau$. Therefore we have displayed them
in four panels, that correspond to the categories A, B, C, D, respectively.

Our model receives no constraints from $B_d \to \mu \mu$ and $B_s \to \mu \mu$ since these decays 
depend on ${\mathcal O}_{10}$, and our charge assignments do not introduce any NP contribution
to this operator. 
The NP contribution will affect $b \to s \nu \nu$ decays, however the current upper limits
\cite{Buras:2014fpa} are 4-5 times larger than the SM predictions, whereas in the region
that is consistent with the neutral meson mixing and global fits for the rare decays, 
the enhancement of this decay rate in our model is not more than 10\%.
See appendix~\ref{section:nunu} for further details.

\subsection{Direct constraints from collider searches for $Z^\prime$}

In figure~\ref{fig:combinedconstraints} we also show the bounds in the 
$g_{Z^\prime}$--$M_{Z^\prime}$ plane from the 95\% upper limits on the 
$\sigma \times {\rm BR}$ for the process $p p \to Z^\prime \to \ell \ell$ \cite{ATLAS:2016cyf,CMS:2016abv}. 
The bounds coming from di-jet final state~\cite{ATLAS:2015nsi,Sirunyan:2016iap} are relatively weaker than those coming from di-leptons, 
hence we neglect the di-jet bounds in our analysis. 
The total cross-section $p p \to Z^\prime \to \ell \ell$ depends not only on 
$M_{Z^\prime}$ and $g_Z^\prime$ but also on the $X$-charges of quarks and leptons, therefore 
the bounds obtained differ for all the nine symmetries in table~\ref{table:selected}.

Note that the experimental limits in \cite{ATLAS:2016cyf,CMS:2016abv} are given in the narrow width 
approximation, whereas the $Z^\prime$ for masses above 2 TeV has broad width for all the symmetry cases
which we have considered. The constraints in the broad width case are generally weaker, therefore
even lighter $Z^\prime$ values than those shown in the figure are allowed.

\section{Predictions for neutrino mixing and collider signals}
\label{section:predictions}

\subsection{Neutrino mass ordering and CP-violating Phases}

The categories A, B, C and D, in table~\ref{table:selected} correspond to different texture-zero
symmetries in the right-handed neutrino mass matrix $M_R$. Through eq.~(\ref{eq:seesaw}),
these predict the light neutrino mass matrix $M_\nu$, which can be related to 
the neutrino masses and mixing parameters via
\begin{equation}
M_{\nu} = - U_{\text{PMNS}} \, M_{\nu}^{\text{diag}} \, U^T_{\text{PMNS}} \;,
\label{equation:minor}
\end{equation} 
where $U_{\rm PMNS}$ is the neutrino mixing matrix parametrized by three mixing angles 
$\theta_{12}$, $\theta_{13}$, $\theta_{23}$, and the Dirac phase $\delta_{cp}$. The diagonal
mass matrix $M_{\nu}^{\text{diag}}= (e^{2i \alpha_1} m_1, e^{2i \alpha_2} m_2, m_3)$ 
incorporates the Majorana phases $\alpha_1$ and $\alpha_2$, in addition to
the magnitudes of the masses, $m_1, m_2$ and $m_3$.
Since the symmetries restrict the form of $M_R$, they are expected to restrict the
possible values of neutrino mixing parameters. While the neutrino mixing angles 
are reasonable well-measured, the values of unknown parameters like 
$\alpha_1, \alpha_2$ and $\delta_{\rm CP}$ may be restricted in each of the scenario. 
In addition, whether the neutrino mass ordering is normal ($m_2^2 < m_3^2$) or
inverted ($m_2^2 > m_3^2$) is also an open question, and some of the symmetries may have
strong preference for one or the other ordering.
The symmetries in table~\ref{table:selected} that yield two-zero textures for $M_R$, viz. 
$L_\mu - L_\tau$, $L_e - 3 L_\mu - L_\tau$, $L_e + 3 L_\mu - L_\tau$ and $L_e - L_\mu \pm 3 L_\tau$ 
have already been explored in this context and the allowed parameter values determined
~\cite{Lavoura:2004tu,2minors,Fritzsch:2011qv,textures,horizontalpaper}.
\begin{figure} [t]
\begin{center}
\includegraphics[width=0.49\textwidth]{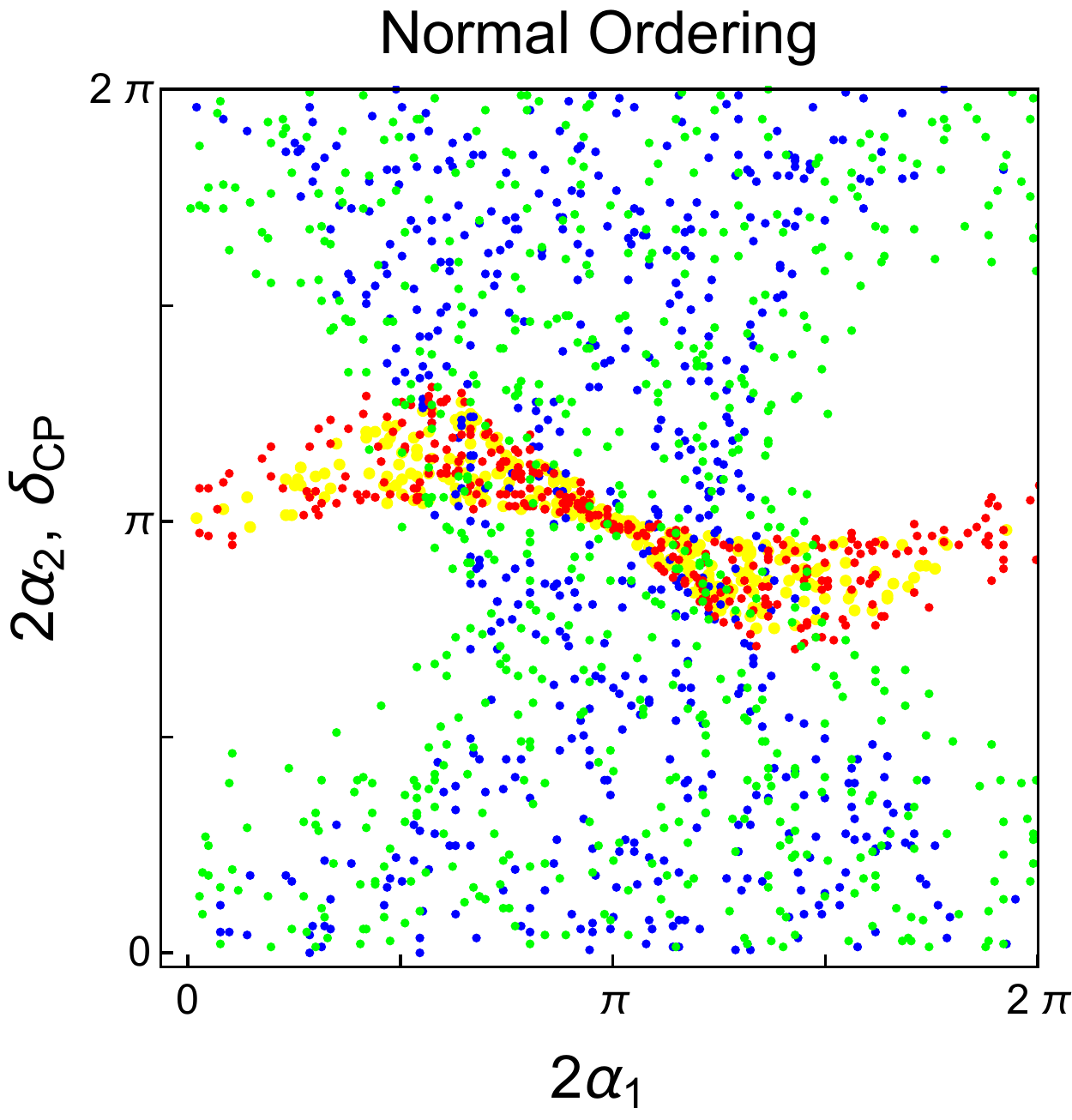}
\includegraphics[width=0.49\textwidth]{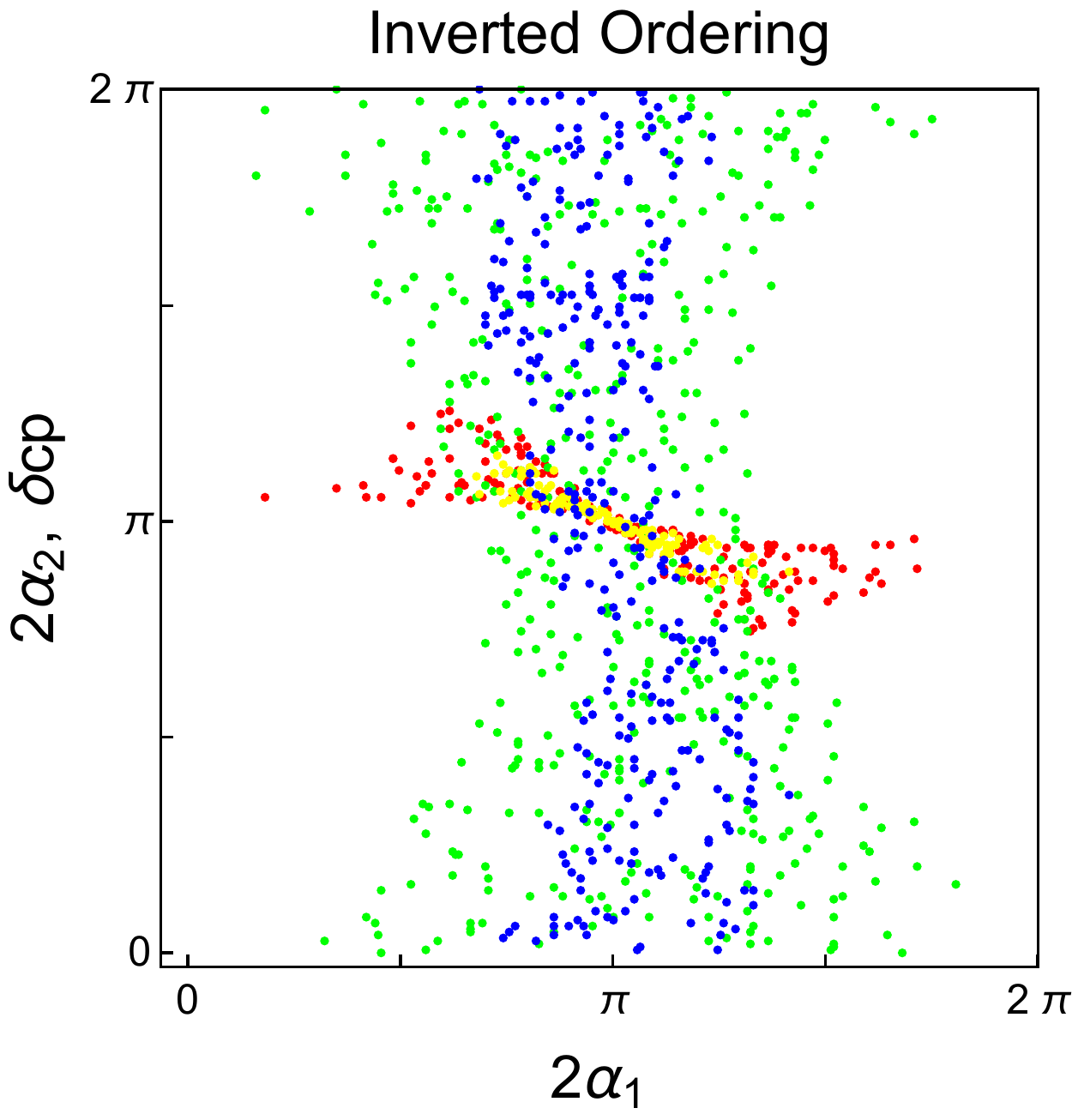}
\caption{The scatter plots of allowed values of the CP phases $\alpha_2$ and $\delta_{\rm CP}$  
with the those of $\alpha_1$. The left (right) panel shows the results for normal (inverted)
mass ordering. The yellow (red) points correspond to $(2\alpha_1, 2\alpha_2)$ values 
for $m_{\rm light} = 0.05 (0.2)$ eV, while 
the blue (green) points correspond to $(2\alpha_1,\delta_{\rm CP})$ values 
for $m_{\rm light} = 0.05 (0.2)$ eV.
}
\label{fig:majorana}
\end{center}
\end{figure}

We exemplify the point in the context of the symmetries that yield one-zero texture for $M_R$,
viz. $L_\mu$ and $L_e-3L_{\mu}+L_{\tau}$. These two also happen to be the ones that are consistent
with all the global fits \cite{globalfit1,globalfit2,globalfit3,globalfit4} to the 
$b \to s \ell\ell$ and $b\to s\gamma$ data to within $1\sigma$. 
Both of these symmetries lead to $[M_R]_{{22}}=0$. Equation~(\ref{eq:seesaw}) then leads to the
condition of one vanishing minor in the $M_{\nu}$ mass matrix \cite{ma}, i.e.
$[M_{\nu}]_{11} [M_{\nu}]_{33} - [M_{\nu}]_{13}^2 = 0$. 
In terms of masses and elements of the $U_{\rm PMNS}$ matrix, 
\begin{eqnarray}
\left( U_{13} U_{32} - U_{12} U_{33}\right)^2 m_2 m_3 e^{2i \alpha_2} 
& =& - \left( U_{12} U_{31} - U_{11} U_{32}\right)^2 m_1 m_2 e^{2i (\alpha_1+ \alpha_2)} \nonumber \\ 
& & -  \left( U_{13} U_{31} - U_{11} U_{33}\right)^2 m_1 m_3 e^{2i \alpha_1} \; , 
\label{eq:one-minor}
\end{eqnarray}
where $U_{ij}$ are elements of the $U_{\text{PMNS}}$ matrix. 
Requiring the neutrino masses and mixings to satisfy the above relation, we show the allowed
values of the CP-violating phases $\alpha_1, \alpha_2$ and $\delta_{\rm CP}$ in
figure~\ref{fig:majorana}, for two fixed values of the lightest neutrino mass $m_{\rm light}$
(i.e. $m_1$ for normal ordering and $m_3$ for inverted ordering). We let the other
neutrino parameters (mixing angles and mass squared differences) to vary within their 
$3\sigma$ ranges~\cite{Capozzi:2016rtj,Esteban:2016qun}. 
The figure shows that the allowed value of $\alpha_2$ with the $L_\mu$ or
$L_e - 3 L_\mu + L_\tau$ symmetry is restricted to be rather close to $\pi/2$.
For lower $m_{\rm light}$ values, $\alpha_2$ is more severely restricted
and for inverted ordering, the value of $\alpha_1$ also 
is restricted to be close to $\pi/2$.

\begin{figure} [t]
\begin{center}
\includegraphics[width=0.49\textwidth]{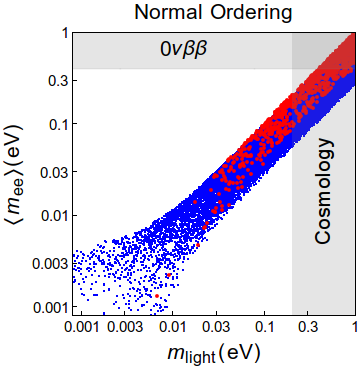}
\includegraphics[width=0.49\textwidth]{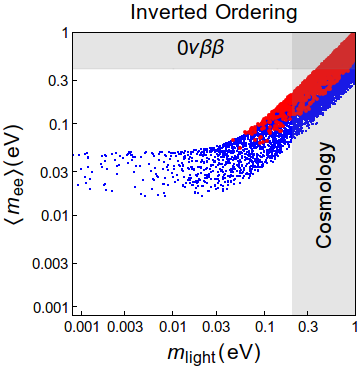}
\caption{The scatter plots of allowed values of $m_{\rm light}$ and $\langle m_{ee} \rangle$.
The red (blue) points correspond to the allowed values with (without) the
symmetry ($L_\mu$ or $L_e - 3 L_\mu + L_\tau$).
The left (right) panel shows the results for normal (inverted) mass ordering.
The regions disallowed by the non-observations of neutrinoless
double beta decay ($0\nu\beta\beta$) and cosmological constraints have also been shown.} 
\label{fig:mee-mlight}
\end{center}
\end{figure}

Another set of predictions may be obtained by relating the lightest neutrino mass 
$m_{\rm light}$ to the effective mass measured by the neutrinoless double beta decay
experiments~\cite{Agostini:2013mzu} if the neutrinos are Majorana, i.e.
\be
\langle m_{ee} \rangle = \left| m_1 e^{2i \alpha_1} \cos^2 \theta_{12} \cos^2 \theta_{13} 
+  m_2 e^{2i \alpha_2} \sin^2 \theta_{12} \cos^2 \theta_{13} +
m_3 e^{- 2 i \delta_{\rm CP}} \sin^2 \theta_{13} \right| \; .
\ee
We show the allowed region (with mixing angles and mass squared differences varied within 
their $3\sigma$ ranges~\cite{Capozzi:2016rtj,Esteban:2016qun}) in the
$m_{\rm light}$--$\langle m_{ee} \rangle$ plane in figure~\ref{fig:mee-mlight}. 
Bounds from the non-observation of neutrinoless double beta decay~\cite{Agostini:2013mzu} 
and conservative limits coming from cosmology ($\sum m_\nu < 0.6$ eV)~\cite{cosmo} have also been shown.
The figure shows that the symmetries $L_\mu$ or  $L_e - 3 L_\mu + L_\tau$ restrict the allowed values of
$m_{\rm light}$ and $\langle m_{ee} \rangle$ significantly in the case of inverted ordering:
$m_{\rm light} \gtrsim 0.045$ eV and $\langle m_{ee} \rangle \gtrsim 0.055$ eV.
With the cosmological bounds on the sum of neutrino masses becoming stronger,
the inverted hierarchy in these scenarios would get strongly disfavoured.

The symmetries in table~\ref{table:selected} that do not lead to a zero-texture in $M_R$,
i.e. $L_e-L_\mu \pm L_\tau$, will not give any predictions for the neutrino mass ordering 
or CP-violating phases; model parameters can always be tuned to satisfy the data.

\subsection{Prospects of detecting $Z^\prime$ at the LHC}
In our model apart from $Z^\prime$, there are additional scalars and three heavy
majorana neutrinos. Note that the parameters in our model have been chosen such that we are in the decoupling
limit, i.e. the additional scalars $H, A, H^\pm, S$ are
too heavy to affect any predictions in the model. The three right handed
neutrinos in our model have masses of the order of a TeV and hence can be 
looked at the collider-based experiments.  
The recent analyses for the 
detection of the heavy right handed neutrinos can be found in~\cite{Antusch:2016ejd}.
We however choose $M_R \gtrsim M_{Z^\prime}/2$, hence 
do not consider the phenomenology of the right handed neutrinos. 

We shall now explore the possibility of a direct detection of the $Z^{\prime}$ gauge boson 
in the 13 TeV LHC run. The cleanest probe for this search is $pp\to Z^{\prime}\to \ell\ell$
~\cite{ATLAS:2016cyf,CMS:2016abv}. 
In such a search, one looks for a peak in the invariant mass spectrum of the dilepton pair. 

\begin{figure} [t]
\begin{center}
\includegraphics[width=0.48\textwidth,height=6.cm]{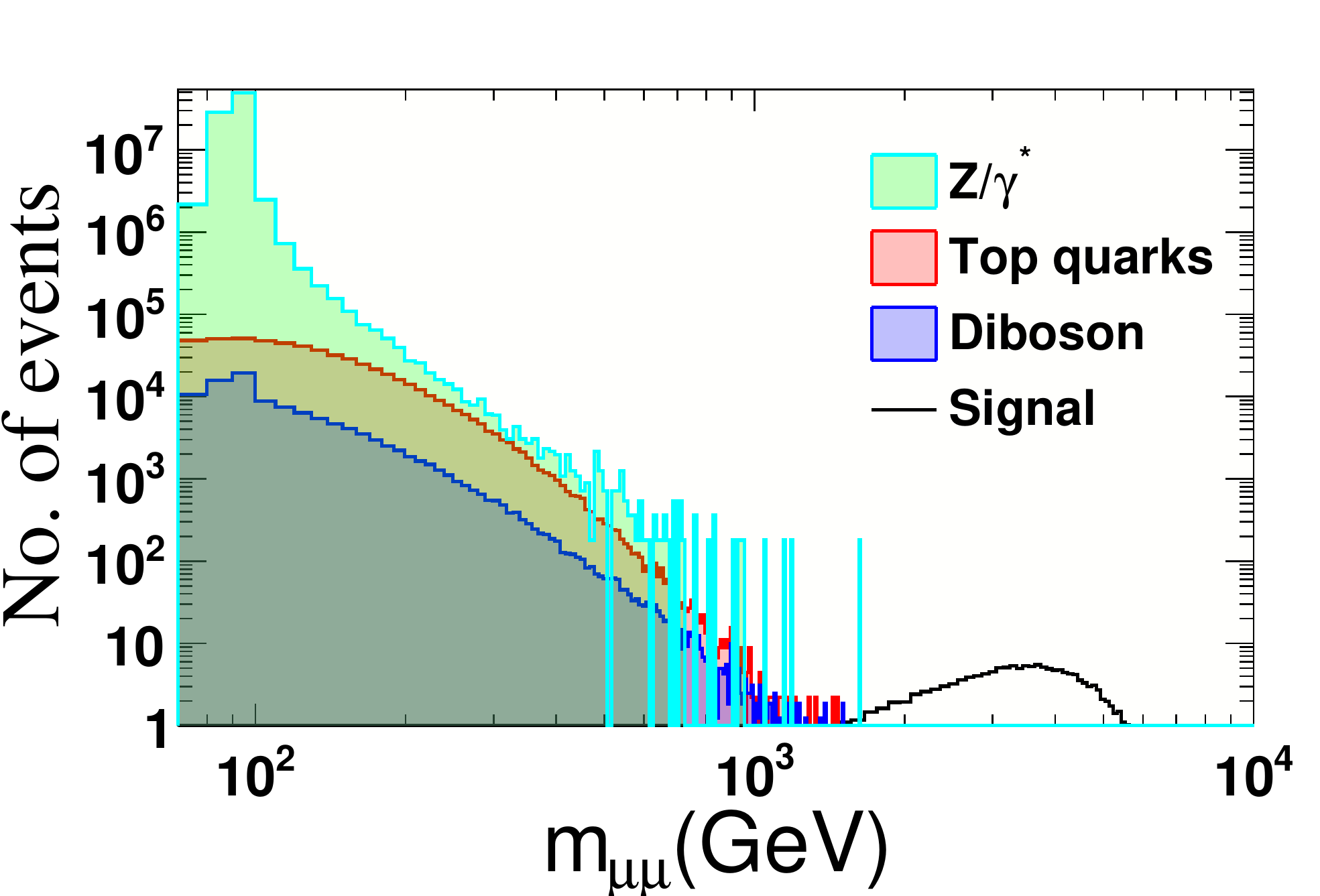}\quad
\includegraphics[width=0.48\textwidth]{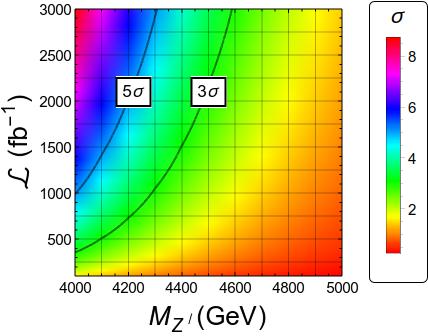}
\caption{\label{fig:4} 
The left panel shows the dimuon invariant mass distribution for the signal
originating from $Z^{\prime}$ (with $M_{Z^\prime} = 4$ TeV and $g_{Z^\prime} = 0.36$)
and the various SM backgrounds at 13 TeV, with $\mathcal{L} = 100$\,fb$^{-1}$.
The right panel shows the discovery significance $S/\sqrt{S+B}$ as a function
of $M_{Z^{\prime}}$ (with $g_{Z^\prime} = 0.36$) and integrated luminosity.
The $5~\sigma$ and $3~\sigma$ contours are also shown explicitly. }
\end{center}
\end{figure}

As an example, we choose the $L_e-3L_{\mu}+L_{\tau}$ symmetry.
We use {\tt FeynRules}~\cite{Alloul:2013bka} to generate the model files and then 
interface the {\tt Madgraph}~\cite{madgraph} output of the model with {\tt PYTHIA 6.4}
~\cite{Sjostrand:2006za} for showering and hadronisation with parton 
distribution function {\tt CTEQ-6}~\cite{Pumplin:2002vw}. The output 
is then fed into {\tt Delphes 3.3}~\cite{Ovyn:2009tx,deFavereau:2013fsa} 
which gives the output in the {\tt ROOT}~\cite{Antcheva:2009zz} format for a semi-realistic 
detector simulation while using the default ATLAS card. In our detector 
analysis jets are constructed from particle flow algorithm using the 
anti-$k_T$ jet algorithm with $R=0.5$ and $p_T^{\text{min}}=50$ GeV. 
We retain events only with a pair of isolated opposite-sign 
muons with highest $p_T$ in each event. Care
has been taken to reject any isolated electron in the event sample.
A rough $p_T$ cut on the muons is set at $p_T^{\mu}>25$ GeV which roughly
matches the ATLAS cuts~\cite{ATLAS:2016cyf}. The dominant SM
background for this di-muon channel comes from the Drell-Yann process. Other
factors contributing to the SM background are diboson and top quarks
in the final state. In the left panel of figure~\ref{fig:4} we show the dimuon
invariant mass distribution of the SM backgrounds as well as the
signal for a fixed benchmark scenario satisfying all the flavour and collider
constraints (see figure~\ref{fig:combinedconstraints}) with 
$M_{Z^\prime}=4$ TeV and $g_{Z^\prime}=0.36$. Although the production
cross section for such a heavy $Z^{\prime}$ gauge boson is small, close
to 1.49 fb, the SM background is also minuscule in that regime. 
Therefore, the $Z^\prime \to \mu\mu$ is a natural probe to look for BSM signals.
We note in passing that a $Z^\prime$ associated with a hard jet in the final state
should increase the signal significance further~\cite{Khachatryan:2016crw}. 
However, we only select events with opposite sign di-muon
pair and a hard jet veto.

To further calculate the reach of the LHC for the $Z^\prime$ discovery via $Z^\prime \to \mu\mu$,
we use a signal specific cut on the dimuon invariant mass $m_{\mu\mu}> 700$ GeV which renders
all the SM backgrounds to be very small whereas the signal hardly gets affected. 
We keep the coupling $g_{Z^\prime}$ fixed at 0.36, and illustrate in the right panel of figure~\ref{fig:4},
the reach of the LHC in the $M_{Z^{\prime}}$-- integrated luminosity ($\mathcal{L}$) plane,
in the form of a density plot of the significance $S/\sqrt{S+B}$ ~\cite{dEnterria:2007iyi}. (Here $S,~B$ are
the number of signal and background events after the cut, respectively.)
The figure indicates that detecting a $Z^{\prime}$ of mass 4000 GeV 
at $3~\sigma$ ($5~\sigma$) significance requires an integrated luminosity close to 
400 fb$^{-1}$ (1000 fb$^{-1}$) in the 13 TeV run of the LHC.

\section{Summary and concluding remarks}
\label{section:conclusions}
 
In this paper, we have looked for a class of models with an additional possible 
$U(1)_X$ symmetry that can explain the flavour anomalies ($R_K$ and $P_5^\prime$) and neutrino 
mixing pattern. The models are built around the phenomenological choice where NP effects 
are dominant only in the $\mathcal{O}_9$ operator, as indicated by the global fits to
the $b \to s$ data. One salient feature of our analysis is that the assignment of $X$-charges 
of fields is done in a bottom-up approach. I.e., we do not start with a pre-visioned symmetry, 
but look for symmetry combinations consistent with both the flavour data and neutrino mixing.

In order to generate neutrino masses through the Type-I seesaw mechanism, we add three
right-handed neutrinos to the SM field content. This also allows us to assign vector-like 
$X$-charges to the SM fermions, so that the anomaly cancellation can be easily achieved.
This choice also makes NP contributions to ${\mathcal O}_{10}$ and ${\mathcal O}_{10}^\prime$ 
vanish. While the different $X$-charge assignments to the SM generations introduce the
desired element of lepton flavour non-universality at tree level, it also introduces the
problem of generating mixings in both quark and lepton sector. This is alleviated by 
adding an additional doublet Higgs $\Phi_1$ that generates the required quark mixing, 
and a scalar $S$ that generates lepton mixing. The choice of rotation matrices 
$V_{u_L} = V_{u_R} =1$, $V_{d_L} = V_{\rm CKM}$ and $V_{d_R} \approx 1$ also ensures that
the NP contribution to ${\mathcal O}_9^\prime$ is negligible. The scalar $S$ also helps 
in avoiding the possible problem of a Goldstone boson appearing from the breaking of 
a symmetry in the doublet Higgs sector. 

Our model is thus rather parsimonious, with the introduction of
only the two additional scalar fields $\Phi_1$ and $S$. The symmetry breaking due to the
vacuum expectation values of these scalars gives mass to the new gauge boson $Z^\prime$,
at the same time keeping its mixing with the SM $Z$ boson under control. 

With the $X$-charges of quark and lepton generations connected through anomaly cancellation,
the $X$-charge assignments may be referred to in terms of the corresponding symmetries
in the lepton sector. We identify those leptonic symmetries that would give rise to the
required structure in the neutrino mass matrix, at the same time are consistent with the
global fits to the $b \to s$ data. We find nine such symmetries, viz. $L_\mu - L_\tau$, $L_\mu$,
$L_e - 3 L_\mu\pm L_\tau$, $L_e + 3 L_\mu - L_\tau$, $L_e - L_\mu \pm 3 L_\tau$, and
$L_e - L_\mu \pm L_\tau$. We find the allowed regions in the $g_{Z^\prime}$--$M_{Z^\prime}$ 
parameter space that satisfy the bounds from neutral meson mixings, rare $B$ decays, and
direct $Z^\prime$ collider searches. 

The lepton symmetries give rise to specific textures in the right-handed neutrino mass
matrix $M_R$, and hence, through seesaw, to patterns in the light neutrino mass matrix.
The consequent neutrino masses and mixing parameters are hence restricted by these
symmetries. In order to exemplify this, we have focussed on the symmetries $L_\mu$ and
$L_e - 3 L_\mu + L_\tau$ that give rise to one zero-texture in $M_R$, and are also the
most favoured symmetries according to all the $b \to s$ global fits.
We have analyzed the correlations among the CP-violating phases $\alpha_1, \alpha_2, \delta_{\rm CP}$,
and also explored the allowed region in the parameter space of the lightest neutrino mass 
$m_{\rm light}$ and the effective neutrino mass $\langle m_{ee} \rangle$ measured in the 
neutrinoless double beta decay.
For $L_e - 3 L_\mu + L_\tau$, we also calculate the reach of the LHC for direct detection 
of $Z^\prime$ through the di-muon channel. We find that discovery of $Z^\prime$ with the
required mass and gauge coupling is possible with a few hundred fb$^{-1}$ integrated 
luminosity at the 13 TeV run.

Note that the parameters in our minimal model have been chosen such that we are in the 
decoupling limit, i.e. the additional scalars $H, A, H^\pm, S$, and the three 
right-handed neutrinos are too heavy to affect any predictions in the model. 
Our model thus does not try to account for the flavour anomalies indicated in the 
semileptonic $b \to c$ decays \cite{Amhis:2014hma}. These anomalies may be addressed in 
the extensions of this minimal model to include non-decoupling scenarios (for example, 
where the charged Higgs is light), or additional charged $W^{\prime \pm}$ gauge bosons. 
While the former scenario needs to satisfy additional constraints from flavour and 
collider data, the latter will need mechanisms for giving masses to the new gauge bosons. 

In this paper we have presented a class of symmetries that are consistent with the current
data, and not applied any aesthetic biases among them. As more data come along, some of these 
symmetries are sure to be further chosen or discarded. We have chosen the symmetries in the 
bottom-up approach, and have not tried to explore their possible origins. 
A curious pattern applicable for some of the symmetries ($L_\mu$, $L_e - 3 L_\mu + L_\tau$ 
and $L_e - L_\mu + L_\tau$) is that the non-universality of $X$-charges is displayed only by
the third generation quarks and the second generation leptons. Such patterns may provide
further hints in the search for the more fundamental theory governing the mass generation
of quarks and leptons.
\acknowledgments
We would like to thank Aoife Bharucha, G. D'Ambrosio, Diptimoy Ghosh, Sandhya Jain, Jacky Kumar, 
Gobinda Majumder, Sreerup Raychaudhuri and Tuhin S. Roy for fruitful discussions. We would 
especially like to thank B. Capdevila and J. Matias for providing us $\chi^2$ values for various 
one-dimensional fits to the $b \to s$ data. This project has received support from the European 
Union’s Horizon 2020 research and innovation programme under the Marie Sklodowska-Curie grant 
agreement Nos. 674896 and 690575.
\appendix
\section{Mass of $Z^\prime$ and $Z$-$Z^\prime$ mixing}
\label{section:zprime}
Our model has three scalar fields: two $SU(2)_L$ doublets $\Phi_1$, $\Phi \equiv \Phi_2$, and 
one singlet $S$. The Lagrangian describing the kinetic terms of the scalar fields is 

 
\begin{eqnarray}
\mathcal{L}^{\text{kin}}_{\text{scalars}} & = &  
\Phi_1^\dagger \left( \overleftarrow{\partial_\mu} 
- i \frac{g_1}{2} W_\mu.\sigma 
- i \frac{g_2}{2}  B_{1\mu} 
- i g_{Z^\prime} X_{\Phi_{1}} B_{2\mu}  \right)  \nonumber \\ 
&&   \left( \overrightarrow{\partial^\mu} 
+ i \frac{g_1}{2} W^\mu.\sigma 
+ i \frac{g_2}{2}  B_1^\mu 
+ i g_{Z^\prime} X_{\Phi_{1}} B_2^\mu  \right) \Phi_1 \nonumber \\
&+&  \Phi_2^\dagger \left( \overleftarrow{\partial_\mu} 
- i \frac{g_1}{2} W_\mu.\sigma 
- i \frac{g_2}{2} B_{1\mu} \right)
\left( \overrightarrow{\partial^\mu} 
+ i \frac{g_1}{2} W^\mu.\sigma 
+ i \frac{g_2}{2} B_1^\mu  \right) \Phi_2 \nonumber  \\
&+& \left(\partial_\mu 
- i g_{Z^\prime} X_{S} B_{2\mu} \right) S^\dagger 
\left({\partial^\mu} 
+ i g_{Z^\prime} X_{S} B_{2}^\mu \right) S  \; .
\end{eqnarray}
With the $X$-charge assignments of scalars as $X_\Phi=0, X_{\Phi_1}=X_S=a$
(see section~\ref{sec:scalars}), the
spontaneous symmetry breaking leads to the following mass term:
\begin{equation}
\mathcal{L}_{{V}}^{\text{mass}} 
=  \frac{1}{2}  
\begin{pmatrix}
W_{3\mu} & B_{1\mu} & B_{2\mu}
\end{pmatrix} 
M^2_{V}
\begin{pmatrix}
W_3^\mu \\ B^{\mu}_1 \\ B^{\mu}_2
\end{pmatrix}\;,
\end{equation}
where
\begin{equation}
M^2_{V} =
\begin{pmatrix}
\frac{1}{4} \,g_1^2 v^2 &&  
- \frac{1}{4} \,g_1\, g_2 v^2  
&& -\frac{1}{2} \,a \, g_1\,g_{Z^\prime} v^2 \cos^2{\beta}\\
-\frac{1}{4} \,g_1 \, g_2 v^2  && 
\frac{1}{4}\, g_2^2 v^2  &&
\frac{1}{2}\, a \, g_2\,g_{Z^\prime} v^2 \cos^2{\beta}\\
-\frac{1}{2}\, a \, g_1\,g_{Z^\prime} v^2 \cos^2{\beta} &&
\,\, \frac{1}{2}\, a \, g_2\, g_{Z^\prime} v^2 \cos^2{\beta} &&
\,\, a^2 \, g_{Z^\prime}^2 \left(v_{S}^2 + v^2 \cos^2{\beta} \right)
\end{pmatrix}\;.
\end{equation}
Since $v \sim$ the electroweak scale, and $v_S \gtrsim $ TeV, we can approximate the
mass eigenstates $\gamma, Z, Z^\prime$ in the limit $v_{S} \gg v$ as \cite{Langacker:2008yv}
\ba
\gamma & = &  \sin{\theta_W} W_{3\mu} + \cos{\theta_W} B_{1\mu} \; ,  \\ 
Z  &  \approx &  \cos{\theta_{Z^\prime}} \left(\cos{\theta_W} W_{3\mu} - \sin{\theta_W} B_{1\mu}\right)
- \sin{\theta_{Z^\prime}} B_{2\mu} \; , \\ 
Z^\prime & \approx &  \sin{\theta_{Z^\prime}} \left(\cos{\theta_W} W_{3\mu} - \sin{\theta_W} B_{1\mu}\right) 
+ \cos{\theta_{Z^\prime}} B_{2\mu} \; ,
\ea
with masses 
\ba
M_\gamma & = & 0 \; , \\
M_Z^2 & \approx & \frac{1}{4} \left( g_1^2 + g_2^2 \right) v^2 - \Delta \; , \\
M_{Z^\prime}^2 & \approx & a^2 g_{Z^\prime}^2 \left( v_{S}^2 + v^2 \cos^2{\beta} \right) + \Delta \; .
\ea
Here $\tan{\theta_W} = g_2/g_1$,
\begin{eqnarray}
\Delta &=& \frac{1}{4}
\frac{ a^2 \left({g_1^2 +g_2^2}\right) g_{Z^\prime}^2 \,v^4 \cos^4{\beta}}
{\left[  a^2 g_{Z^\prime}^2\left(v_{S}^2 + v^2 \cos^2{\beta}\right) 
- \frac{1}{4}\left(g_1^2 + g_2^2 \right) v^2   \right]} \; , \\
\sin{\theta_{Z^\prime}} &\simeq & -\frac{1}{{2}}
\frac{ a \sqrt{g_1^2 +g_2^2} \,g_{Z^\prime} \,v^2 \cos^2{\beta}}
{\left[ a^2 g_{Z^\prime}^2\left( v_{S}^2 + v^2 \cos^2{\beta}\right) 
- \frac{1}{4}\left(g_1^2 + g_2^2 \right) v^2   \right]} \; .
\end{eqnarray}
Note that since $v \ll v_S$, we have $\Delta \sim v^2 (g_1^2 + g_2^2) (v/ v_S)^2 \cos^4\beta \ll v^2$,
and therefore 
\be
M_{Z^\prime} \approx a g_{Z^\prime} v_S
\ee
Also, the $Z$--$Z^\prime$ mixing angle $\theta_{Z^\prime}$ is given by
\be
\sin\theta_{Z^\prime} \approx \frac{\sqrt{g_1^2 + g_2^2} v^2 \cos^2 \beta}{2 a g_{Z^\prime} v_S^2} 
= \frac{M_Z}{M_{Z^\prime}} \frac{v}{v_S} \cos^2\beta \; .
\ee 
Thus, the $Z$--$Z^\prime$ mixing is automatically suppressed: $\theta_{Z^\prime} \sim {\mathcal O}(10^{-3})$. 
Therefore, it would not affect our model.

\section{Controlling flavour changing neutral currents mediated by scalars}
\label{section:FCNC}

When the singlet $S$ is heavy and effectively decoupled, the scalar doublets $\Phi_1$ and $\Phi_2$ 
can be parameterized as~\cite{branco}
\begin{equation}
H_1 = \cos{\beta} \, \Phi_1 + \sin{\beta} \, \Phi_2 \;, \quad  
H_2 = -\sin{\beta} \, \Phi_1 + \cos{\beta} \, \Phi_2 \;.
\end{equation}
where
\ba
H_1 & = & \begin{pmatrix}
0 \\ 
\frac{1}{\sqrt{2}} \big[ h \, \sin(\alpha-\beta) - H \, \cos(\alpha-\beta) + i\, G^0 + v \big]
\end{pmatrix}\;, \quad \nonumber \\
H_2 & = & \begin{pmatrix}
H^+ \\ 
\frac{- 1}{\sqrt{2}} \big[ h \,  \cos(\alpha-\beta) + H \, \sin(\alpha-\beta)  - i \, A \big]
\end{pmatrix} \;,
\ea
such that only the combination $H_1$ gets a vacuum expectation value.
Here $h$ is the SM-like Higgs with mass equal to 125 GeV, and $H$, $A$ and $H^\pm$ 
are the heavy Higgs, psuedo-scalar Higgs and the charged Higgs, respectively. 
The Lagrangian in eqn.~(\ref{eq:Lyuk2}) expressed in terms of $H_1$ and $H_2$ is
\begin{eqnarray}
\mathcal{L}_{\rm Yuk} &=&  
\overline{Q^{\rm f}_{L}} \, \bigg[ 
\left( {\mathcal Y}_1^{u} \cos\beta + {\mathcal Y}^u \sin\beta \right) H_1^c 
-\left( {\mathcal Y}_1^{u} \sin\beta - {\mathcal Y}^u \cos\beta \right) H_2^c 
 \bigg]  u^{\rm f}_{R} \nonumber \\
&& + \,\, \overline{Q^{\rm f}_{L}} \,  \bigg[
 \left( {\mathcal Y}_1^{d} \cos\beta + {\mathcal Y}^d \sin\beta \right) H_1 
-\left( {\mathcal Y}_1^{d} \sin\beta - {\mathcal Y}^d \cos\beta \right) H_2  
 \bigg] d^{\rm f}_{R} \;.
\label{eq:Lyuk3}
\end{eqnarray}
This Lagrangian can be expanded to obtain the flavour changing neutral current (FCNC)
interactions mediated by the Higgs bosons.
Since the up-type quark mass matrix has been chosen to be flavour diagonal
(see section~\ref{section:ckm}), there are no tree-level FCNC's in the up sector. 
The tree-level FCNC's of $d$-type quarks hence can be written as
\begin{equation}
\mathcal{L}^{\text{FCNC}}_{d} = \frac{1}{\sqrt{2}}\overline{d_L} 
V^\dagger_{\rm CKM}\left[
\bigg( {\mathcal Y}_1^{d} \sin\beta - {\mathcal Y}^d \cos\beta \bigg)
\bigg( H \sin{\left(\alpha - \beta\right)} + h \cos{\left(\alpha - \beta\right)} - i A \bigg) 
\right] d_R \; .
\end{equation}
From the above Lagrangian, it can be seen that the FCNC contributions of $H$ and $A$ have
opposite signs and hence they tend to cancel if $\alpha-\beta \approx \frac{\pi}{2}$ 
and $M_A \approx M_H$. The FCNC contribution of the light Higgs $h$ also vanishes for 
$\alpha-\beta \approx \frac{\pi}{2}$. Such limits naturally appear in the decoupling scenarios 
for two-Higgs doublet models, and can be easily incorporated by the suitable choice of parameters 
in the eq.~(\ref{equation:Vpot}). The scalar spectrum in our model 
is $M_h \ll M_H, M_A, M_{H^\pm} \ll M_S$. 

Note that though the charged Higgs $H^\pm$ will not contribute to tree-level FCNC, it will have
contributions through the penguin and box diagrams. In the decoupling scenario, such contributions 
would be miniscule and may be ignored.

\section{Enhancement or suppression of $b\to s\nu \nu$}
\label{section:nunu}
\begin{figure}[b]
\begin{center}
\includegraphics[totalheight=7cm]{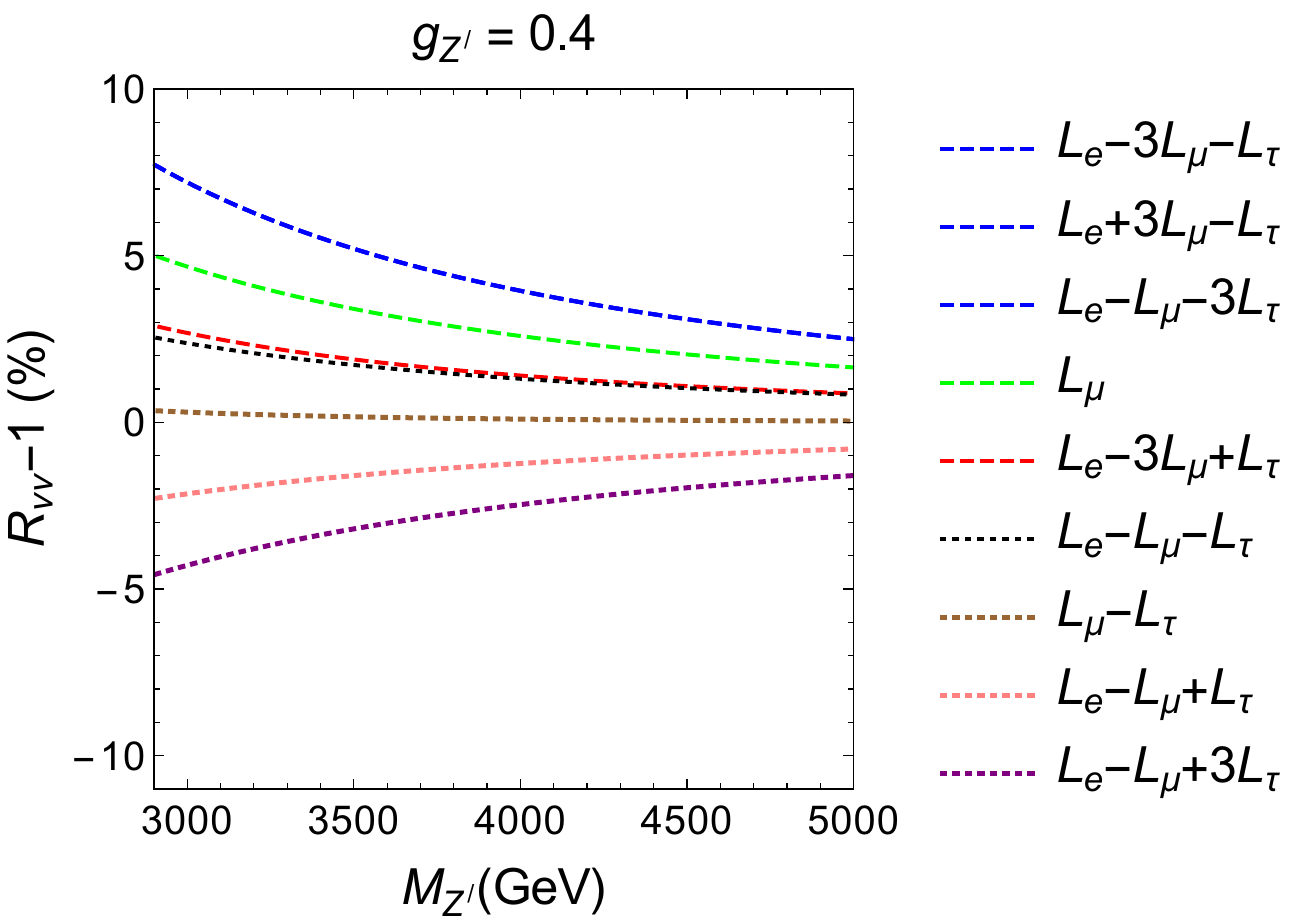}
\caption{\label{fig:nunu} 
Predictions for $R_{\nu\nu}$ with different symmetries from table~\ref{table:selected}.}
\end{center}
\end{figure}

The effective Hamiltonian for $b \to s \nu_\ell \nu_\ell$ in SM is~\cite{Buras_review,Buras:2014fpa}
\begin{eqnarray}
\mathcal{H}_{\text{eff}}^{\text{SM}}& =& -\dfrac{4 G_F}{\sqrt{2}}\, \frac{\alpha_e}{4 \pi} \, V_{tb} V_{ts}^* \, C^{\text{SM}}_{L}
\big[\overline{s_L}\gamma_\mu \,b_L \big]\, 
\big[\overline{\nu_\ell}\gamma_\mu (1-\gamma_5)\nu_\ell\big] \;,
\end{eqnarray}
where $C^{\text{SM}}_{L} = -X_t/s_W^2$, with $X_t=1.469\pm 0.017$~\cite{Buras:2014fpa}.
The $Z^\prime$ mediation also generates the contribution to the same operator. The 
combined SM and NP effect is
\begin{eqnarray}
\mathcal{H}_{\text{eff}}^{\text{tot}}& =& -\dfrac{4 G_F}{\sqrt{2}}\, \frac{\alpha}{4 \pi} \, V_{tb} V_{ts}^* \, 
( C^{\text{SM}}_{L} + C^{\text{NP},\ell}_{L} )
\big[\overline{s_L}\gamma_\mu \,b_L \big]\, 
\big[\overline{\nu_\ell}\gamma_\mu (1-\gamma_5)\nu_\ell\big] \;,
\end{eqnarray}
with $C^{\text{NP},\ell}_{L} = (x_1-x_3) \pi y_\ell g^2_{Z^\prime}/(\sqrt{2} M^2_{Z^\prime}  G_F \alpha_e)$
The right handed current operator contributions are small (see arguments leading
to eq.~(\ref{equation:C9prime})) and are neglected. NP can enhance the rate of an individual
lepton channel $b \to s \nu_\ell \nu_\ell$ if $(x_1-x_3) y_\ell < 0$. 
In experiments, the branching ratios and the decay widths corresponding to $b \to s \nu_\ell \nu_\ell$ 
has to summed over all the three generations of neutrinos. 
We consider the quantity $R_{\nu\nu}$ which gives us a measure of NP effects
\begin{equation}
R_{\nu\nu} = \frac{| C^{\text{SM}}_{L}+C^{\text{NP},e}_{L}|^2 + | C^{\text{SM}}_{L}+C^{\text{NP},\mu}_{L}|^2
 + | C^{\text{SM}}_{L}+C^{\text{NP},\tau}_{L}|^2}{3 \,| C^{\text{SM}}_{L}|^2}
 \end{equation}
The enhancement or suppression of the branching ratio crucially depends on the combined effects 
of $(x_1-x_3)y_\ell$ for the three generations.

In figure~\ref{fig:nunu} we show the value of $R_{\nu\nu}$ as a function of $M_{Z^\prime}$
for all the symmetries in table~\ref{table:selected}, where
the coupling has been fixed to $g_{Z^\prime} = 0.4$. 
It is observed that the net increment is not more than 10\% for all symmetries.
(Note that for some symmetries, the lower values of masses may not be allowed, as shown in 
figure~\ref{fig:combinedconstraints}, in which case the deviation would be further reduced.)
The enhancement and suppression is thus too small for the current experiments to be
sensitive to -- The current bounds on $\text{BR}(B \to K^{(*)} \nu \overline{\nu})$
are 4--5 times higher than the SM prediction~\cite{Buras:2014fpa}, while Belle2 experiment is 
expected to reach a sensitivity close $30\%$ from SM by 2023~\cite{Aushev:2010bq}.

\end{document}